\documentclass{wsm}
\usepackage{amssymb,epsfig,bm}
\newcommand{\ind}[2]{^{#1}_{\mbox{\scriptsize #2}}}
\newcommand{\al}[2]{\alpha\ind{#1}{#2}}
\newcommand{\tal}[2]{\widetilde{\alpha}\ind{#1}{#2}}
\newcommand{\hal}[2]{\widehat{\alpha}\ind{#1}{#2}}
\newcommand{\bt}[2]{\beta\ind{#1}{#2}}
\newcommand{\ro}[1]{\rho^{(#1)}}
\newcommand{\Ro}[1]{{\cal R}^{(#1)}(\sigma)}
\newcommand{\DRo}[2]{\Delta{\cal R}^{(#1)}(#2)}
\newcommand{\QbQ}{$\mbox{Q}\overline{\mbox{Q}}$ }

\newcommand{\RTau}[1]{R_{\tau,\mbox{\tiny #1}}}

\def\DCI{D^{\mbox{\tiny CI}}}
\def\LQCD{$\Lambda_{\mbox{\tiny QCD}}$ }
\def\sfa{\mbox{\sf A}}
\def\nf{n_{\mbox{\scriptsize f}}}
\def\nfs{n_{\mbox{\tiny f}}}
\def\Nc{N_{\mbox{\scriptsize c}}}
\def\Ncs{N_{\mbox{\tiny c}}}
\def\bz{\beta_0}
\def\bu{\beta_1}
\def\Vud{V_{\mbox{\scriptsize ud}}}
\def\Sew{S_{\mbox{\tiny EW}}}
\def\dQCD{\delta_{\mbox{\tiny QCD}}}
\def\Reg{{\mbox{\scriptsize Reg}}}
\def\MSbar{$\overline{\mbox{MS}}$ }
\def\KL{K\"all\'en--Lehmann }

\def\Ctwo{\mbox{C}^{(2)}(y)}
\def\Atwo{\mbox{A}^{(2)}(y)}
\def\Cthree{\mbox{C}^{(3)}(y)}
\def\Athree{\mbox{A}^{(3)}(y)}

\renewcommand{\Re}{\mbox{\rm Re} \;}

\newlength{\intwidth}
\newcommand{\pvint}[2]{
\settowidth{\intwidth}{$\displaystyle\int\limits_{#1}^{#2}$}
\mbox{\hbox to 0pt{\hbox
to\intwidth{{\hfill{\raisebox{0.45mm}{\scriptsize$-$}}\hfill}}\hss}$\displaystyle\int\limits_{#1}^{#2}$}
}

\begin{document}

\markboth{}{}
\catchline{Vol.~18, No.~30 (2003) 5475--5519}{}{}{}{}

\title{ANALYTIC$\,$ INVARIANT$\,$ CHARGE$\,$ IN$\,$ QCD}

\author{\footnotesize A.V.~NESTERENKO}

\address{Centre de Physique Theorique de l'\'Ecole Polytechnique \\
91128 Palaiseau Cedex, France\footnote{Unit\'e Mixte de Recherche
du CNRS (UMR 7644)}\\
nesterav@cpht.polytechnique.fr}
\address{Bogoliubov Laboratory of Theoretical Physics,
Joint Institute for Nuclear Research \\
Dubna 141980, Russian Federation\\
nesterav@thsun1.jinr.ru}

\maketitle

% Preprint CPHT-RR 028.0603
\pub{Received 16 June 2003}{}

\begin{abstract}
This paper gives an overview of recently developed model for the QCD
analytic invariant charge. Its underlying idea is to bring the
analyticity condition, which follows from the general principles of
local Quantum Field Theory, in perturbative approach to
renormalization group (RG) method. The concrete realization of the
latter consists in explicit imposition of analyticity requirement on
the perturbative expansion of $\beta$~function for the strong running
coupling, with subsequent solution of the corresponding RG equation.
In turn, this allows one to avoid the known difficulties originated in
perturbative approximation of the RG functions. Ultimately, the
proposed approach results in qualitatively new properties of the QCD
invariant charge. The latter enables one to describe a wide range of
the strong interaction processes both of perturbative and
intrinsically nonperturbative nature.

\keywords{Nonperturbative QCD; analytic approach;
renormalization group; strong running coupling.}

\vskip2.5mm
\noindent
PACS numbers: 12.38.Lg, 11.10.Hi, 11.55.Fv, 12.38.Aw
\end{abstract}

% SECTION 1

\section{Introduction}

     Theoretical analysis of strong interaction processes rests on
the quantum non-Abelian gauge field theory, namely, Quantum
Chromodynamics (QCD). This theory is dealing with quarks and gluons.
The former are the constituents of hadrons, while the latter are
quanta of a massless gauge vector field, which provides interaction
between quarks. Gluons and quarks carry a specific quantum number,
which plays a key role in hadron physics. It was first introduced in
the mid 1960s in Ref.~\refcite{Color1}, and later it was
named\cite{Color2} ``color''. The comprehensive theoretical and
experimental investigations revealed that the strong interaction
possesses two distinctive features. First, the interaction between
quarks inside the hadrons weakens with increasing the characteristic
energy of a process. In other words,  the invariant
charge\footnote{Sometimes invariant charge $\alpha(q^2)= \bar{g}^2
(q^2)/(4 \pi)$ is also called the (strong) running coupling of~QCD.}
vanishes at large momenta transferred, representing the so-called
asymptotic freedom of the theory. Second, free quarks and gluons have
not been observed, at least until the present time. Here we are
dealing with the color confinement.

     These two phenomena are related to hadron dynamics at different
energy scales. The asymptotic freedom takes place at large values of
energies, or in the so-called ultraviolet (UV) region. It corresponds
to small ($r\lesssim0.1\,$fm) quark separations. On the contrary, the
color confinement is related to the low-energy, or infrared (IR)
domain, that corresponds to large quark separations, namely,
$r\gtrsim1.0\,$fm.

     Theoretical description of the asymptotic freedom was first
performed in the mid 1970s in Refs.~\refcite{AF}. These papers gave
rise to wide  employment of perturbative calculations in Quantum
Chromodynamics. Indeed, if the value of invariant charge is small,
then one can parameterize unknown quantities by perturbative power
series in the running coupling. On the one hand, this drastically
simplifies  analysis of hadron interactions in the UV region. On the
other hand, the hadron dynamics at low energies, in particular, the
quark confinement, entirely remains beyond the scope of perturbation
theory. As a rule, description of the QCD experimental data in the IR
region requires invoking other nonperturbative approaches, for
instance, phenomenological potential models
(Refs.~\refcite{EQRM}--\refcite{Melles}),  sum rules method
(Refs.~\refcite{SVZ}--\refcite{CK}),   string
models,\cite{Nest,Solov} bag models,\cite{Bags}  variational
perturbation theory  (Refs.~\refcite{VPT1}--\refcite{VPT4}), and
lattice simulations (Refs.~\refcite{Bali5}--\refcite{JQCD}).

     The renormalization group (RG) method plays a fundamental role
in the framework of Quantum Field Theory (QFT) and its applications.
It was developed almost half a century ago in  Refs.~\refcite{RG1}
and~\refcite{RG2} (see also Ref.~\refcite{BgSh}). Nowadays,
theoretical analysis of hadron dynamics relies, in main part,
upon the RG method. Meanwhile, the construction of exact solutions to
the renormalization group equation, which is a rather nontrivial
mathematical problem itself, is still far from being feasible.
Usually, in order to describe the strong interaction processes in
asymptotical UV region, one applies the RG method together with
perturbative calculations. In this case, the {\it a priori} unknown
renormalization group functions are parameterized by the power series
in the QCD running coupling. Ultimately, this leads to approximate
solutions of RG equation, which are commonly used  for quantitative
analysis of experimental data. However, employment of perturbation
theory in the framework of renormalization group method  generates
unphysical singularities of such solutions. The latter are in
contradiction with the basic principles of local Quantum Field
Theory. For example, solution to RG equation for invariant charge
gains unphysical singularities due to perturbative approximation of
the  $\beta$~function. At the one-loop level there is the so-called
Landau pole, and account of the higher-loop corrections just
introduces additional singularities of the cut type into expression
for the running coupling (see~\ref{Sect:RCPert}).

     An effective way to overcome such difficulties was
proposed\cite{Redm} in the late 1950s by P.J.~Redmond. It consists in
invoking into  consideration the analyticity requirement, which
follows from the general principles of local QFT. This prescription
became the underlying idea of the so-called analytic approach to QFT,
which was formulated in the framework of Quantum Electrodynamics
(QED) by N.N.~Bogoliubov,  A.A.~Logunov, and D.V.~Shirkov.\cite{BLS}
Really, the QED running coupling is proportional to the transverse
part of the dressed photon propagator. Therefore, in accordance with
the basic principles of local QFT,\cite{BgSh} the spectral
representation of the \KL type holds for it. The latter implies the
definite analytic properties in $q^2$~variable for the invariant
charge. Namely, it should be the analytic function in the complex
$q^2$--plane with the only cut along the negative\footnote{A metric
with signature $(-1, 1, 1, 1)$ is used, so that positive $q^2$
corresponds to a spacelike momentum transfer.} semiaxis of
real~$q^2$.

     In the general case  the QCD invariant charge is defined as a
product of propagators and a vertex function (see, e.g.,
Ref.~\refcite{Nar}). Therefore, one might pose a question concerning
the analytic properties of this quantity. This matter has been
examined in Ref.~\refcite{GSh}, and it was shown therein that in this
case the integral representation of the \KL type holds for the
running coupling,~too. Proceeding from these motivations, the
analytic approach was lately extended\cite{APTMod} to Quantum
Chromodynamics by D.V.~Shirkov and I.L.~Solovtsov,  and applied to
the ``analytization'' of perturbative series for  the QCD
observables\footnote{This method has been named the analytic
perturbation theory (APT) thereafter.} (see reviews~\refcite{APTRev}
and Refs.~\refcite{APTTL}--\refcite{APTSR} for details). The term
``analytization'' means here the restoring of the correct analytic
properties in the $q^2$ variable of a quantity under consideration by
making use of the \KL integral representation
\begin{equation}
\label{DefAn}
\Bigl\{\sfa(q^2)\Bigr\}_{\mbox{$\!$\scriptsize an}} =
\int\limits_{0}^{\infty} \!\frac{\varrho(\sigma)}{\sigma+q^2}\,
d \sigma.
\end{equation}
In this equation the spectral function $\varrho(\sigma)$ is
determined by the initial  (perturbative) expression for a quantity
in hand
\begin{equation}
\varrho(\sigma) = \frac{1}{2\pi i}\, \lim_{\varepsilon \to 0_{+}}
\Bigl[\sfa(-\sigma-i\varepsilon)-\sfa(-\sigma+i\varepsilon)\Bigr],
\qquad \sigma > 0.
\end{equation}
It is worth mentioning also the appealing features of the analytic
approach to Quantum Field Theory, namely, the absence of unphysical
singularities and fairly well higher loop and scheme stability of
outcoming results.

     The description of hadron interaction at low energies remains a
long--standing challenge of Quantum Chromodynamics. Some insight into
many basic problems of elementary particle physics can be gained form
constructing the models, which adequately account both the
perturbative and intrinsically nonperturbative traits of hadron
dynamics. Certainly,  such studies are useful for further development
of the strong interaction theory. Besides, the description and
interpretation of many experimental data can be performed only by
invoking such models. Thus, the results obtained in this field could
expand our scope of knowledge concerning the nature of fundamental
interactions.

     This paper overviews a new model for the QCD analytic invariant
charge. The model is formulated in detail with a specific attention
to the properties of the analytic running coupling and to
verification of the model's self-consistency. The principally new way
of involving the analyticity condition into the RG formalism is a
distinguishing feature of this model. Namely, the analyticity
requirement is imposed on the perturbative expansion of the
renormalization group $\beta$~function. Eventually it leads to a
number of profitable traits of this running coupling, that enables
one to describe a wide range of the strong interaction processes both
of perturbative and intrinsically nonperturbative nature. It is worth
noting also that this model for the strong running coupling has no
adjustable parameters. Therefore, similarly to perturbative approach,
\LQCD remains the basic characterizing parameter of the theory.

     The layout of the review is as follows.

     In Section~\ref{Sect:Model} the model for QCD analytic invariant
charge is discussed. The key point here is the restoring of the
correct analytic properties of perturbative expansion for the
$\beta$~function, and the subsequent solution of the corresponding RG
equation. The explicit expression for the one-loop analytic invariant
charge (AIC) is obtained. It is proved to have qualitatively new
properties. In particular, the analytic invariant charge has no
unphysical singularities, and it incorporates UV asymptotic freedom
with IR enhancement in  a single expression. The AIC is found to be
in agreement with the results of various nonperturbative studies of
hadron dynamics in the infrared domain. For example, this strong
running coupling explicitly reproduces the recently  discovered
symmetry\cite{Schrempp01} related to the size distribution of
instantons.  The consistency of the considered model with  general
definition of the QCD invariant charge is proved. The analytic
running coupling is also represented in explicitly renorminvariant
form, and the properties of the relevant $\beta$~function are
studied. A number of symmetry relations for the analytic invariant
charge and the corresponding $\beta$~function, which establish a link
between the ultraviolet and infrared domains, are derived. The
different ways of incorporating the analyticity requirement into the
RG formalism are discussed.

     Section~\ref{Sect:HL} is devoted to investigation of the
analytic invariant  charge at the higher-loop levels. The integral
representation for this running coupling is derived herein. Similarly
to the one-loop level, analytic invariant charge is shown to have no
unphysical singularities. The developed model possesses a good higher
loop and scheme stability. The investigation of the corresponding
$\beta$~function revealed that AIC has the universal asymptotics in
both the ultraviolet and infrared regions, irrespective of the loop
level. Further, the model in hand is extended to the timelike domain.
The difference between values of the analytic running coupling in
respective spacelike and timelike regions is found to be considerable
for intermediate and low energies. Apparently, this circumstance must
be taken into account when one handles the experimental data. The
result obtained in this section confirms the hypothesis due to
Schwinger concerning the $\beta$~function proportionality to the
relevant spectral density.

     The applications of the analytic invariant charge to study of a
number of strong interaction processes are gathered in
Section~\ref{Sect:Applic}. In particular, the static quark--antiquark
potential constructed by making use of the AIC is proved to be
confining at large distances. At the same time, at small distances it
has the standard behavior originated in asymptotic freedom. Further,
the developed approach is applied to description of gluon condensate,
inclusive $\tau$~lepton decay, and electron--positron annihilation
into hadrons. The congruity of estimated values of the scale
parameter \LQCD testifies that the analytic invariant charge
substantially incorporates, in a consistent way, both perturbative
and intrinsically nonperturbative aspects of Quantum Chromodynamics.

     In the Conclusions (Section~\ref{Sect:Concl}) the properties of the
analytic invariant charge are summarized and the basic results are
formulated in a compact way.

     The auxiliary materials are collected in the appendixes.
In~\ref{Sect:RCPert} the perturbative solutions to the RG equation
for QCD running coupling are derived at different loop levels.
In~\ref{Sect:LambW} the Lambert~$W$ function is briefly described.
The function $N(a)$, which essentially simplifies investigation of
the $\beta$~function corresponding to the one-loop AIC, is defined
and studied here. \ref{Sect:SF} contains explicit expressions for the
spectral functions used in construction of the analytic invariant
charge. In~\ref{Sect:VrExp} the coefficients specifying the expansion
of the quark--antiquark potential are presented.

% SECTION 2

\section{The QCD Analytic Invariant Charge}
\label{Sect:Model}

\subsection{Formulation of the model}
\label{Sect:ModelDef}

     First of all, let us consider the renormalization group equation
for the strong running coupling (see, e.g., Refs.~\refcite{BgSh} and
\refcite{Ynd}):
\begin{equation}
\label{RGGen}
\frac{d\,\ln \bigl[g^2(\mu)\bigr]}{d\,\ln\mu^2} =
\beta\Bigl(g(\mu)\Bigr).
\end{equation}
In the framework of perturbative approach, the $\beta$ function on
the right-hand side of this equation can be represented as a power
series
\begin{equation}
\label{BetaPert}
\beta\Bigl(g(\mu)\Bigr) = - \left\{
\beta_{0}\left[\frac{g^2(\mu)}{16 \pi^2}\right] +
\beta_{1}\left[\frac{g^2(\mu)}{16 \pi^2}\right]^2 + \ldots \right\},
\end{equation}
where $\beta_{0} = 11 - 2 \nf / 3,\,$ $\beta_{1}=102 - 38 \nf / 3$,
and $\nf$ is the number of active quarks. Introducing the standard
notations  $\al{}{s}(\mu^2)= g^2(\mu)/(4\pi)$ and $\tal{}{}(\mu^2)
= \alpha(\mu^2)\, \beta_{0}/(4\pi)$, one can reduce the RG
equation~(\ref{RGGen}) at the $\ell$-loop level to the form
\begin{equation}
\label{RGPert}
\frac{d\,\ln\bigl[\tal{(\ell)}{s}(\mu^2)\bigr]}{d\,\ln \mu^2} = -
\sum_{j=0}^{\ell-1} \beta_j
\left[\frac{\tal{(\ell)}{s}(\mu^2)}{\beta_{0}}\right]^{j+1}.
\end{equation}
It is well-known that the solution to this equation has unphysical
singularities at any loop level. So, there is the Landau pole at the
one-loop level, and account of the higher loop corrections just
introduces the additional singularities of the cut type into
expression for the invariant charge.\footnote{Solutions to
equation~(\ref{RGPert}) at different loop levels are considered in
detail in~\ref{Sect:RCPert}.} However, the fundamental
principles of local Quantum Field Theory require the invariant charge
$\alpha(q^2)$ to have a definite analytic properties in the $q^2$
variable. Namely, there must be the only cut along the negative
semiaxis of real~$q^2$ (see Ref.~\refcite{APTMod}).

     The solution to the renormalization group equation~(\ref{RGGen})
gains unphysical singularities due to perturbative approximation of
the $\beta$~function~(\ref{BetaPert}).  Such a representation of the
right-hand side of the RG equation proves to be justified in the
ultraviolet region, where the asymptotic freedom takes place.
However, the approximation~(\ref{BetaPert}) does not ultimately lead
to the correct analytic properties of the QCD invariant charge. In
fact, since the strong running coupling  $\alpha(q^2)$ is of a fixed
sign, the condition that invariant charge has the only left cut
in~$q^2$ means that the right-hand side of the RG
equation~(\ref{RGGen}), as a function of $q^2$, has no singularities
in the region of positive~$q^2$. Apparently, the perturbative
expansion~(\ref{BetaPert}) violates this condition.

     In the framework of developed model\cite{PRD1,PRD2} the
analyticity requirement is imposed on the $\beta$~function
perturbative expansion for restoring its correct analytic properties.
This leads to the following RG equation for the QCD analytic
invariant charge:
\begin{equation}
\label{AnRGEq}
\frac{d\,\ln\bigl[\tal{(\ell)}{an}(\mu^2)\bigr]}{d\,\ln \mu^2} =
- \left\{\sum_{j=0}^{\ell-1} B_{j}
\Bigl[\tal{(\ell)}{s}(\mu^2)\Bigr]^{j+1}\right\}_{\!\mbox{\scriptsize an}},
\qquad B_{j} = \frac{\beta_{j}}{\beta_{0}^{j+1}}.
\end{equation}
In this equation $\al{(\ell)}{an}(\mu^2)$ denotes the $\ell$-loop
analytic invariant charge, $\al{(\ell)}{s}(\mu^2)$ is the
perturbative running coupling at the $\ell$-loop level, and the
braces $\bigl\{\,\ldots\,\bigr\}\ind{}{an}$ mean the analytization of
the expression within them by making use of the \KL spectral
integral~(\ref{DefAn}) (see also Refs.~\refcite{France01}
and~\refcite{ConfV}).

\subsection{The one-loop analytic invariant charge}

     Since the proposed model rests on invoking the analyticity
requirement into the renormalization group formalism, one may
anticipate that the strong running coupling acquires qualitatively
new properties. Let us proceed to construction of solution to the RG
equation~(\ref{AnRGEq}), restricting ourselves to the leading loop
approximation at this stage.

     At the one-loop level the renormalization group
equation~(\ref{AnRGEq}) takes the form\cite{PRD1,PRD2}
\begin{equation}
\label{RGAn1L}
\frac{d\,\ln \bigl[\tal{(1)}{an}(\mu^2)\bigr]}{d\,\ln\mu^2} =
- \Bigl\{\tal{(1)}{s}(\mu^2)\Bigr\}_{\mbox{$\!$\scriptsize an}}.
\end{equation}
Here $\al{(1)}{an}(\mu^2)$ is the one-loop analytic invariant charge
and $\al{(1)}{s}(\mu^2)$ denotes the perturbative running coupling at
the one-loop level. Taking into account Eq.~(\ref{DefAn}), one can
represent the right-hand side of the RG equation~(\ref{RGAn1L}) as
follows
\begin{equation}
\label{RGEqnAn1}
\frac{d\,\ln \bigl[\tal{(1)}{an}(\mu^2)\bigr]}{d\,\ln\mu^2} =
- \int\limits_{0}^{\infty} \frac{\Ro{1}}{\sigma+\mu^2}\, d\sigma,
\end{equation}
where $\Ro{1} = [\ln^2 (\sigma /\Lambda^2) + \pi^2]^{-1}$. Upon the
integration with respect to $\sigma$ and the introduction of the
dimensionless variable $\nu^2 = \mu^2 / \Lambda^2$,
Eq.~(\ref{RGEqnAn1}) acquires the form
\begin{equation}
\frac{d\,\ln \bigl[\tal{(1)}{an}(\nu^2\Lambda^2)\bigr]}{d\,\ln\nu^2} =
- \left[\frac{1}{\ln \nu^2} + \frac{1}{1 - \nu^2}\right].
\end{equation}
Further, integrating this equation with respect to $\nu^2$ in finite
terms, one gets
\begin{equation}
\label{AIC1LNorm}
\frac{\tal{(1)}{an}(q^2)}{\tal{(1)}{an}(q_0^2)} =
\frac{z-1}{z\,\ln z} \, \frac{z_0\,\ln z_0}{z_0-1},
\end{equation}
where $z = q^2/\Lambda^2$ and $z_0 = q_0^2/\Lambda^2$. Thus, the QCD
analytic invariant charge at the one-loop level has the
form\footnote{Equation~(\ref{AIC1L}) has been derived by  employing
the condition $\al{(1)}{an}(q^2) \to \al{(1)}{s}(q^2)$, when $q^2 \to
\infty$ (see also discussion in Subsection~\ref{Sect:AICHL}).}
\begin{equation}
\label{AIC1L}
\al{(1)}{an}(q^2) = \frac{4 \pi}{\bz} \, \frac{z - 1}{z \, \ln z},
\qquad z = \frac{q^2}{\Lambda^2}.
\end{equation}
It is worth mentioning also that for this strong running coupling
the \KL representation
\begin{equation}
\label{AIC1LKL}
\al{(1)}{an}(q^2) = \frac{4 \pi}{\bz} \int\limits_{0}^{\infty}
\frac{\ro{1}(\sigma)}{\sigma+z} \, d\sigma
\end{equation}
holds, where $\ro{1}(\sigma)$ is the one-loop spectral density
\begin{equation}
\label{SpDns1L}
\ro{1}(\sigma) =
\left(1+\frac{1}{\sigma}\right)\frac{1}{\ln^2\sigma+\pi^2}.
\end{equation}

\begin{figure}
\centerline{\psfig{file=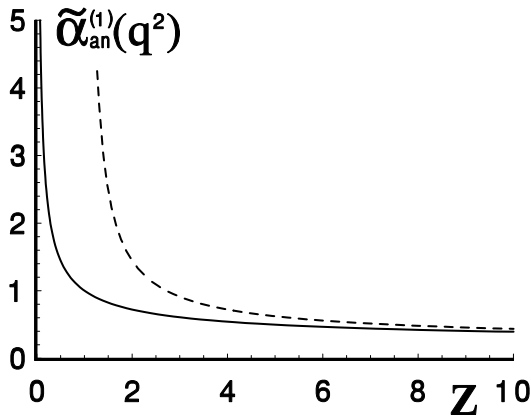,width=70mm}}
\vspace*{8pt}
\caption{The one-loop analytic invariant charge $\tal{(1)}{an}(q^2)$
defined by Eq.~(\protect\ref{AIC1L}) (solid curve) and its
perturbative analog $\tal{(1)}{s}(q^2)$ (dashed curve), $z = q^2 /
\Lambda^2$.}
\label{Plot:AIC1L}
\end{figure}

     Figure~\ref{Plot:AIC1L} depicts the analytic invariant charge
$\tal{(1)}{an}(q^2) = \al{(1)}{an}(q^2)\,\bz/(4 \pi)$ together with
the one-loop perturbative running coupling $\tal{(1)}{s}(q^2) =
1/\ln(q^2/\Lambda^2)$. The obtained solution to the RG
equation~(\ref{AIC1L}) possesses a number of profitable features.
First of all, the integral representation of the \KL
type~(\ref{AIC1LKL}) implies that the analytic invariant charge
$\al{(1)}{an}(q^2)$ has the only cut\footnote{The function $a(z) =
(z-1) / (z\,\ln z)$ has a smooth behavior in the vicinity of the
point $z=1$, namely $a(1 + \varepsilon) \simeq 1 - \varepsilon /2 +
{\cal O}(\varepsilon^2),\;\; \varepsilon \to 0$.} along the negative
semiaxis of real~$q^2$. Thus, the prescription proposed ultimately
leads to elimination of the Landau pole in  a  multiplicative way.
One has to note that the analytic  invariant charge~(\ref{AIC1L})
incorporates  the ultraviolet asymptotic  freedom with the infrared
enhancement in  a single expression. In particular, this trait will
play an essential role in applications of the elaborated model, see
Section~\ref{Sect:Applic} further. It is worth emphasizing also that
the developed model contains no adjustable parameters, i.e.,
similarly to perturbative approach, \LQCD remains the basic
characterizing parameter of the theory.

     A crucial insight into the nonperturbative aspects of the
low-energy hadron dynamics can be provided by the relevant lattice
simulations. One should mention here a recent lattice investigation
of the topological structure of the SU(3) vacuum by the UKQCD
Collaboration.\cite{UKQCD} It was revealed therein that the size
distribution of instantons~$D(\rho)$ has a conspicuous peak shape,
and it is severely suppressed at large scales~$\rho$. In turn, this
is compatible neither with perturbative results, nor with the
freezing of the strong running coupling to some constant value at
large distances.\cite{UKQCD,Schrempp99}  Remarkably, the analytic
invariant charge~(\ref{AIC1L}) is proved to reproduce explicitly a
special kind of symmetry related to this issue\cite{Schrempp01} (see
Subsection~\ref{Sect:AIC1L} for details). Meantime, the strong
running coupling itself can also be studied on the lattice. In
particular, a recent investigation of its low-energy behavior by
ALPHA Collaboration\cite{ALPHA} testify to the infrared enhancement
of the QCD invariant charge. The study of the Schwinger--Dyson
equations is another source of the nonperturbative information
concerning the strong interaction issues. It is worth noting here
that the behavior of the QCD invariant charge, similar to
Eq.~(\ref{AIC1L}), is found to be in agreement with the solution of
the Schwinger--Dyson equations (see papers~\refcite{AlekArbu,Alek98}
and references therein for details).

\medskip

     In general, the QCD invariant charge is defined as the product
of the corresponding Green functions and vertexes. For example, in
the transverse gauge the following definition
\begin{equation}
\label{InvChrgDef}
\alpha(q^2) = \alpha(\mu^2)\, G(q^2)\, g^2(q^2)
\end{equation}
takes place (see, e.g., Ref.~\refcite{Nar}). In this equation
$\alpha(q^2)$ is the QCD running coupling, $\alpha(\mu^2)$ denotes
its value at a normalization point $\mu^2$, $G(q^2)$ and $g(q^2)$ are
dimensionless gluon and ghost propagators, respectively. It is
essential to note here that the Green functions possess the correct
analytic properties in the $q^2$ variable\footnote{Namely, there is
the only cut along the negative semiaxis of real~$q^2$.} before the
RG summation. Indeed, in massless theory the Green function can be
written as the perturbative power series in $\alpha(\mu^2)$ with the
coefficients being a polynomial in $\ln (q^2/\mu^2)$. Obviously, in
this case the integral representation of the \KL type must hold for
any finite order of perturbative expansion for the Green
function.\footnote{One should note that  the relevant spectral
function here can not be directly identified with the
spectral density.} However, the RG summation leads to violation of
such a property of the Green functions (e.g., in the simplest
one-loop case the Landau pole appears). As it has already been
mentioned, the analytic approach to QCD seeks to restore the correct
analytic properties of the relevant quantities. It turns out that for
the consistency of involving the analyticity requirement with the
definition of the invariant charge~(\ref{InvChrgDef}), one has to
apply the analytization procedure~(\ref{DefAn}) to the logarithmic
derivative of the Green function $d\,\ln G(q^2)/d\,\ln q^2$. Really,
if the spectral function of the \KL representation for the Green
function $G(q^2)$ is of a fixed sign (that is true for the leading
orders of perturbation theory), then $G(q^2)$ has no zeros in the
complex $q^2$--plane. Hence, the derivative $d\,\ln G(q^2)/d\,\ln
q^2$ can also be represented as the spectral integral of the \KL
type  (cf.~equations (\ref{RGEqnAn1}) and (\ref{AIC1LKL}) for the
one-loop  level, and Eqs.~(\ref{RGEqnAnHL1}) and (\ref{AICHLKL}) for
the higher loop levels).

     In the framework of perturbative approach the Green functions
considered above have the following form: $g(q^2) =
(1/\ln\,z)^{d_{g}}$ and $G(q^2) = (1/\ln\,z)^{d_{G}}$, where
$z=q^2/\Lambda^2$, $d_{g}$ and $d_{G}$ denote the corresponding
anomalous dimensions. For the latter the relationship $2 d_{g} +
d_{G} = 1$ holds, that plays the key role in
definition~(\ref{InvChrgDef}). Applying the analytization procedure
to these functions in the way specified above, one arrives at the
following result: $g\ind{}{an}(q^2) = \left[(z-1)/(z\,\ln z)\right]
^{d_{g}}$ and $G\ind{}{an}(q^2) = \left[(z-1)/(z\,\ln z)\right]
^{d_{G}}$. Therefore, in this case invoking the analyticity condition
does not affect\footnote{It is interesting to note that the similar
situation takes place in the nonperturbative $a$-expansion
method also.\cite{VPT1,VPT3}} the definition of the invariant
charge~(\ref{InvChrgDef}). Thus, we infer that the analytic running
coupling~(\ref{AIC1L})  is consistent with the general definition of
the QCD invariant charge (see also discussion of this matter in
Ref.~\refcite{MPLA2}).

\subsection{Properties of the one-loop analytic invariant charge}
\label{Sect:AIC1L}

     Let us address now the renorminvariance of the analytic running
coupling~(\ref{AIC1L}). In general, any expression for the invariant
charge makes sense only if the relevant definition of the parameter
\LQCD is provided. Otherwise, the running  coupling may not be a
renorminvariant quantity at all. In fact, the renormalization
invariance of a solution to the RG equation~(\ref{RGPert}) is ensured
by a certain relation between the parameter \LQCD and the value of
the running coupling $\alpha(\mu^2)$ at a reference point~$\mu^2$
(see Eqs.~(\ref{LDef1L}) and  (\ref{LDef2L}) also). However, the way
of introduction of the parameter $\Lambda_{\mbox{\tiny QCD}}$,
considered in the previous section, blurs such a relation.
Nevertheless, in the framework of the model in hand, the latter can
be recovered by solving the equation\footnote{Equation~(\ref{RGInv})
is identical to the normalization condition $\bar{g}^2(1,g) = g^2$
for the QCD invariant charge $\bar{g}^2(q^2/\mu^2,g)/(4\pi) =
\alpha(q^2)$ (see, e.g., Ref.~\refcite{SlFad}).}
\begin{equation}
\label{RGInv}
\al{(1)}{an}(q^2) \biggr|_{q^2=\mu^2} = \alpha(\mu^2)
\end{equation}
with respect to parameter~$\Lambda$. In~\ref{Sect:LambW} equations of
this type are studied in detail. Thus, the solution to
Eq.~(\ref{RGInv}) can be explicitly written down in terms of the
Lambert~$W$ function which is defined by the relation\footnote{The
properties of the Lambert~$W$ function~(\ref{WDef}) and the function
$N(a)$ (\ref{NDef}) are described in detail in~\ref{Sect:LambW}.}
\begin{equation}
\label{WDef}
W_{k}(x)\,\exp\Bigl[W_{k}(x)\Bigr] = x.
\end{equation}
For our purposes it is convenient to introduce the function~$N(a)$
\begin{equation}
\label{NDef}
N(a) = \left\{
\begin{array}{ll}
N_{0}(a),        & 0 < a \le 1, \\[2.5mm]
N_{-1}(a), \quad & 1 < a,       \\
\end{array} \right.\qquad
N_{k}(a) = -a\, W_{k}\!
\left[-\frac{1}{a}\,\exp\!\left(-\frac{1}{a}\right)\right].
\end{equation}
In Eqs.~(\ref{WDef}) and (\ref{NDef}) $k$ denotes the branch index of
the Lambert~$W$ function. It is straightforward to verify that the
solution to Eq.~(\ref{RGInv}) has the form
\begin{equation}
\label{LDef}
\Lambda^2 = \mu^2\, N\!\left[\frac{\beta_0}{4 \pi} \alpha(\mu^2)\right].
\end{equation}
Let us note here that the right-hand side of this equation tends to
the one-loop perturbative form when $\mu^2 \to \infty$ (see
Eqs.~(\ref{LDef1L})  and (\ref{NSeries1}) also):
\begin{equation}
\Lambda\ind{2}{s} = \mu^2\,\exp\!\left[- \frac{4 \pi}{\bz}
\frac{1}{\alpha(\mu^2)}\right].
\end{equation}
Thus, the renorminvariant expression for the analytic running
coupling~(\ref{AIC1L}) is the following:
\begin{equation}
\al{(1)}{an}(q^2) = \frac{4 \pi}{\bz} \, \frac{z - 1}{z \, \ln z},
\qquad z = \frac{q^2}{\Lambda^2}, \qquad
\Lambda^2 = \mu^2\, N\!\left[\frac{\beta_0}{4 \pi}
\alpha(\mu^2)\right],
\end{equation}
where the function $N(a)$ is defined in Eq.~(\ref{NDef}) (see
also Ref.~\refcite{MPLA1} for details).

\medskip

     In the framework of the elaborated model the analyticity
requirement is imposed on the $\beta$~function perturbative expansion
for restoring its correct analytic properties. Apparently, the form
of  the $\beta$~function in this case is affected by the
analytization procedure. Thus, it is of a particular interest to
study both the $\beta$~function itself and its properties within the
approach  developed.

     In general, the $\beta$~function corresponding to the one-loop
analytic invariant charge~(\ref{AIC1L}) is defined by\footnote{In
Ref.~\refcite{MPLA1} the definition $\beta(a) = d\,a(\mu^2)/d\,\ln
\mu^2$ was used.}
\begin{equation}
\label{BetaDef}
\bt{(1)}{an}(a) = \frac{d \,\ln a(\mu^2)}{d \,\ln\mu^2},
\qquad a(\mu^2) = \tal{(1)}{an}(\mu^2).
\end{equation}
The function~$N(a)$ (see Eq.~(\ref{NDef})) enables one to obtain the
explicit expression for the right-hand side of Eq.~(\ref{BetaDef}):
\begin{equation}
\label{BetaAn}
\bt{(1)}{an}(a) = \frac{1 - N(a)/a}{\ln\left[N(a)\right]}.
\end{equation}
In order to examine the asymptotics of the
$\beta$~function~(\ref{BetaAn}) it is worth employing the
Eqs.~(\ref{NSeries1})--(\ref{NSeries2}). It turns out that for small
values of the running coupling~$a(\mu^2)$ equation~(\ref{BetaAn})
coincides with the well-known perturbative result
\begin{equation}
\label{BetaOrig}
\bt{(1)}{an}(a) = - a +
{\cal O}\!\left[\frac{1}{a}\exp\!\left(-\frac{1}{a}\right)\right],
\qquad a \to 0_{+}.
\end{equation}
The second term in this equation implies the intrinsically
nonperturbative nature of the $\beta$~function (\ref{BetaAn}).  Then,
for large values of~$a(\mu^2)$ one has
\begin{equation}
\label{BetaInf}
\bt{(1)}{an}(a) = - 1 +
{\cal O}\!\left[\frac{\ln(\ln a)}{(\ln a)^2}\right],
\qquad a \to \infty.
\end{equation}
Such a behavior of the $\beta$~function leads to the infrared
enhancement of the QCD invariant charge, namely (up to a logarithmic
factor) $\tal{}{}(q^2) \simeq \Lambda^2/q^2$, when $q^2 \to 0$. It is
worth mentioning also that
\begin{equation}
\bt{(1)}{an}(1 + \varepsilon) = - \frac{1}{2} - \frac{\varepsilon}{6}
+ {\cal O}(\varepsilon^2), \qquad \varepsilon \to 0,
\end{equation}
i.e., the function~(\ref{BetaAn}) has a smooth behavior at the point
of interchange between the real branches of the Lambert~$W$ function.
Moreover, $\bt{(1)}{an}(a) \leq 0$ for any non-negative value of
$a(\mu^2)$, that ensures the asymptotic freedom of the theory.

\begin{figure}
\centerline{\psfig{file=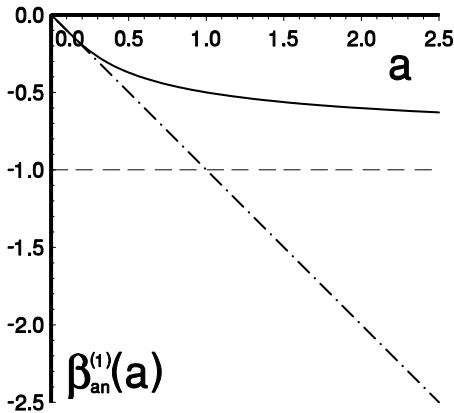,width=60mm}}
\vspace*{8pt}
\caption{The $\beta$~function (\protect\ref{BetaAn}) (solid curve)
corresponding to the one-loop analytic invariant
charge~(\protect\ref{AIC1L}). The relevant perturbative
result~$\bt{(1)}{s}(a)=-a$ is shown by the dot-dashed line.}
\label{Plot:Beta1L}
\end{figure}

     Figure~\ref{Plot:Beta1L} presents the $\beta$~function
(\ref{BetaAn}) together with its perturbative analog
$\bt{(1)}{s}(a)=-a$, which corresponds to the one-loop running
coupling $\tal{(1)}{s}(q^2) = 1/\ln(q^2/\Lambda^2)$. In particular,
this plot visualizes the reproduction  of perturbative
limit~(\ref{BetaOrig}) within the model elaborated.

     Thus, the proposed way of involving the analyticity requirement
into RG formalism eventually leads to the qualitatively new features
of the $\beta$~function~(\ref{BetaAn}). Namely, it incorporates the
asymptotics related to both the UV asymptotic
freedom~(\ref{BetaOrig}) and the IR enhancement~(\ref{BetaInf}) of
the QCD invariant charge. In particular, such behavior of the
$\beta$~function corresponds to the so-called  ``V--scheme'' (see
Refs.~\refcite{KKO}--\refcite{Kiselev} and~\refcite{Peter}), which is
widely used in studies of the quark confinement.

\medskip

     Let us address now the symmetries of the analytic invariant
charge $\al{(1)}{an}(q^2)$ and the corresponding
$\beta$~function~$\bt{(1)}{an}(a)$, which relate these quantities in
the ultraviolet and infrared domains. Obviously, such relations are
decisive, since they provide us with insight into the intrinsically
nonperturbative features of the strong interaction. The symmetry of
this kind for the analytic invariant charge~(\ref{AIC1L}) was found
in Refs.~\refcite{Schrempp01} and~\refcite{MPLA2}:
\begin{equation}
\label{AICSymm}
\al{(1)}{an}(q^2) = \frac{\Lambda^2}{q^2}\,
\al{(1)}{an}\!\left(\frac{\Lambda^4}{q^2}\right).
\end{equation}
It is interesting that there are similar symmetry relations for the
$\beta$~function~(\ref{BetaAn}) also. In
particular,\footnote{Equation~(\ref{BetaSymm}) can be deduced from
Eq.~(\ref{AICSymm}).}
\begin{equation}
\label{BetaSymm}
\beta\Bigl(\tal{(1)}{an}(q^2)\Bigr) +
\beta\Bigl(\tal{(1)}{an}(\Lambda^4/q^2)\Bigr) = - 1
\end{equation}
and
\begin{equation}
\label{AICBetaSymm}
\frac{\beta\left(\tal{(1)}{an}(q^2)\right)}{\tal{(1)}{an}(q^2)} +
\frac{\beta\left(\tal{(1)}{an}(\Lambda^4/q^2)\right)}
{\tal{(1)}{an}(\Lambda^4/q^2)} = - 1,
\end{equation}
where $\beta (a(\mu^2))=d\ln a(\mu^2)/d\ln \mu^2$. In fact, there is
a perturbative analog of Eq.~(\ref{AICBetaSymm}), namely
$\beta(\tal{(1)}{s}(q^2))/\tal{(1)}{s}(q^2) = -1$, which holds
for all non-negative~$q^2$.

\medskip

     From the general viewpoint, one might anticipate that these
symmetries, interrelating the hadron dynamics at short and long
distances, could be revealed also in some intrinsically
nonperturbative strong interaction processes. It is  worth mentioning
here the recently discovered conformal inversion symmetry related to
the size distribution of instantons.\cite{Schrempp01} At the
classical level, the whole instanton sector is proved to be invariant
under the four-dimensional conformal transformations.\cite{JNR} In
particular, the coordinate inversion
\begin{equation}
\label{CoordInv}
x_\mu \longrightarrow x_\mu^\prime = \frac{\rho_0^2}{x^2}\, x_\mu
\end{equation}
transforms the SU(2) Yang--Mills instanton
solution\cite{Inst,tHooft} of size~$\rho$ in singular gauge to the
anti-instanton solution of size $\rho_0^2/\rho$ in regular gauge.
However, at the quantum level it is rather difficult to expect that
the instanton sector is still invariant under the conformal group
transformations since the scale invariance is broken via the
regularization and renormalization. Nevertheless,  there are
evidences that the invariance under the coordinate
inversion~(\ref{CoordInv}), which has a number of important physical
implications, exists, in a certain form, at the quantum level, too.
Thus, the essential result in this field was obtained in
Ref.~\refcite{Schrempp01}, where it was observed that the
high-quality lattice simulation data\cite{UKQCD,Schrempp99} for the
quantity
\begin{equation}
\DCI(\rho) = D(\rho) \left(\frac{\rho}{\rho_0}\right)^{\!\Ncs/6},
\qquad D(\rho) = b \left[\frac{2 \pi}{\alpha(\rho)}\right]^{2 \Ncs}
\exp\!\left[- \frac{2 \pi}{\alpha(\rho)}\right]
\end{equation}
appear to satisfy with high precision the conformal inversion
symmetry $\DCI(\rho) = \DCI(\rho_0^2/\rho)$. Here $D(\rho)$ is the
size distribution of instantons,\cite{ISD} $\rho$ is the radius of an
instanton, $\alpha(\rho)$ denotes the strong running coupling,
$\rho_0$ is the peak position of~$\DCI(\rho)$, $b$ is a constant
factor depending on the gauge group, and $\Nc$ is the number of
colors. This symmetry of $\DCI(\rho)$ is evidently originated in the
mentioned above conformal invariance of the instanton sector at the
classical level, and it turns out to be at  most weakly broken at the
quantum level.

     Apparently, the conformal inversion symmetry $\DCI(\rho) =
\DCI(\rho_0^2/\rho)$ imposes a certain constraint onto the QCD
invariant charge, leading to an important relation between the
properties of the $\alpha(\rho)$ at small and large distances. In
fact, one can derive the expression for the nonperturbative strong
running coupling by employing the symmetry of this kind. Following
this way, the analytic invariant charge~(\ref{AIC1L}) has recently
been  rediscovered and proved\cite{Schrempp01} to reproduce
explicitly the conformal inversion symmetry of $\DCI(\rho)$. This
remarkable result opens up an alluring opportunity to build up  a new
general method of consistent description of the both perturbative and
intrinsically nonperturbative aspects of hadron dynamics towards the
infrared regime.

\subsection{Other models}

     Several decades ago, when the analytic approach to Quantum
Electrodynamics was developing, it was noted\cite{KFF} that there is
no unique way of incorporating the analyticity condition into the RG
formalism. There was also an attempt\cite{Arb59} to resolve this
ambiguity by involving an additional condition, in particular, by
making use of the equations of motion. Following this way the photon
propagator has been derived,\cite{Arb59} the unphysical singularity
being removed in a multiplicative way (i.e., similarly to
Eq.~(\ref{AIC1L})). Furthermore, the presence of nonperturbative
terms both in the running coupling itself~(\ref{AIC1L}) and in the
corresponding spectral density~(\ref{SpDns1L}) is another common
feature of the model in hand and the approach considered in
Ref.~\refcite{Arb59}  (see also discussion of this issue in
Refs.~\refcite{PRD2,MPLA2} and~\refcite{MPLA1}).

     The identical situation arises in Quantum Chromodynamics,
namely, different models for  the strong running coupling can be
proposed in the framework of the analytic approach to QCD. As it was
mentioned above, our model\cite{PRD1,PRD2} is based on recovering the
correct analytic properties of the perturbative expansion of the RG
$\beta$~function~(\ref{AnRGEq}). Meanwhile, in the original model due
to Shirkov and Solovtsov\cite{APTMod} the analytization
procedure~(\ref{DefAn}) is directly applied to the perturbative
invariant charge (see also reviews~\refcite{APTRev}). At the one-loop
level it gives
\begin{equation}
\label{ARCShSol}
\al{(1)}{{\tiny SS}}(q^2) = \frac{4 \pi}{\bz} \left[\frac{1}{\ln z} +
\frac{1}{1-z}\right], \qquad z=\frac{q^2}{\Lambda^2}.
\end{equation}
From the general point of view, one can anticipate the different
properties of these two running couplings. On the one hand, both
these models have no unphysical singularities and reproduce
perturbative limit in the ultraviolet region. On the other hand, there
is a difference  between them in the infrared domain. Namely, the
analytic invariant charge~(\ref{AIC1L}) possesses the IR~enhancement,
while the running coupling~(\ref{ARCShSol}) freezes to the universal
constant value $\al{}{{\tiny SS}}(0)=4 \pi / \bz$. One has to
emphasize here that both these models have no adjustable parameters,
so they are the ``minimal'' ones in this sense. A brief discussion of
this issue can also be found in Refs.~\refcite{PRD2,MPLA2}
and~\refcite{Sh3}.

     It is worth mentioning some other models for the strong running
coupling within the analytic approach to QCD. Let us start with the
model developed by Alekseev and Arbuzov.\cite{AlekArbu,Alek98} By making
use of a special solution to the Schwinger--Dyson equations for the
gluon propagator, these authors proposed the following expression for
the QCD running coupling:
\begin{equation}
\label{ARCAlArb}
\al{(1)}{{\tiny AA}}(q^2) = \frac{4 \pi}{\bz} \left[
\frac{1}{\ln z} + \frac{1}{1-z} + \frac{c}{z} +
\frac{1-c}{z+m_g^2/\Lambda^2}\right], \qquad z=\frac{q^2}{\Lambda^2},
\end{equation}
where $m_g$ is the gluon mass and $c$ denotes a dimensionless
parameter fixed by the phenomenological value of the gluon
condensate. Likewise the analytic  invariant charge~(\ref{AIC1L}),
the running coupling~(\ref{ARCAlArb}) has enhancement in the infrared
domain.

     It is interesting to note here that another similar model for
the QCD invariant charge
\begin{equation}
\label{ARCLatt}
\al{(1)}{{\tiny Latt}}(q^2) = \frac{4 \pi}{\bz} \left[
\frac{1}{\ln z} + \frac{1}{1-z} + \frac{\eta}{z}\right],
\qquad z=\frac{q^2}{\Lambda^2}
\end{equation}
has been put forward in Ref.~\refcite{FrLatt1} for the analysis of
the lattice simulation data on the low--energy behavior of the QCD
running coupling.\cite{FrLatt1,FrLatt2} The  model~(\ref{ARCLatt})
also possesses the infrared enhancement.

     By making use of a certain phenomenological reasoning, Webber
suggested the strong running coupling of  the following
form\cite{Webber}
\begin{equation}
\al{(1)}{{\tiny W}}(q^2) = \frac{4 \pi}{\bz} \left[
\frac{1}{\ln z} + \frac{1}{1-z}\,\frac{z+b}{1+b}
\left(\frac{1+c}{z+c}\right)^{\!p}\,
\right], \qquad z=\frac{q^2}{\Lambda^2},
\end{equation}
with a specific choice of the parameters $b=1/4$, $c=4$, and $p=4$.
On the contrary, this model has infrared finite value
$\al{(1)}{{\tiny W}}(0) \simeq 2 \pi / \bz$.

     There is also a generalization of the analytic invariant
charge~(\ref{AIC1L}). Thus, the authors of Ref.~\refcite{SPPW} proposed
the IR enhanced model for the QCD running coupling
\begin{equation}
\al{(1)}{{\tiny SPPW}}(q^2) = \left[
\frac{1}{\al{(1)}{{\tiny SPPW}}(\Lambda^2)} +
\frac{\bz}{4 \pi} \int\limits_{0}^{\infty}
\frac{(z-1)\, z^p}{(\sigma+z-i\varepsilon)(\sigma+1)(1+z^p)}
\,d \sigma \right]^{-1},
\end{equation}
where $z=q^2/\Lambda^2$ and $0 < p \leq 1$. This formula coincides
with Eq.~(\ref{AIC1L}) when~$p=1$.

     Several phenomenological models for the QCD running coupling
have been proposed in Ref.~\refcite{KrPi01}. It should be mentioned
that the ideas, similar to that of the analytic perturbation
theory,\cite{APTMod,APTRev} were also used in analysis of the
electron--positron annihilation into hadrons\cite{MI} and
investigation of the inclusive $\tau$ lepton decay.\cite{Ioffe2} The
thorough study of the power corrections to the strong running
coupling was performed in Refs.~\refcite{Grunberg}
and~\refcite{Fischer}. There is also a number of methods of the RG
improvement of perturbative series for the QCD observables (see,
e.g., Ref.~\refcite{RGImpr}).

% SECTION 3

\section{Higher Loop Levels}
\label{Sect:HL}

\subsection{Analytic invariant charge at the higher loop levels}
\label{Sect:AICHL}

     As it has been shown in Section~\ref{Sect:Model}, the one-loop
analytic invariant charge possesses a number of appealing features.
The absence of unphysical singularities and incorporation of the
ultraviolet asymptotic freedom with the infrared enhancement in a
single expression are the most remarkable ones. Apparently, the
question of a primary interest here is whether these results are
affected by the higher loop corrections or not. So, let us proceed to
the study of the  analytic running coupling at the higher loop
levels.

     In accordance with the model proposed (see
Subsection~\ref{Sect:ModelDef}), the QCD analytic invariant charge
$\al{(\ell)}{an}(q^2)$ at the $\ell$-loop  level is the solution to
the renormalization group equation\cite{PRD2}
\begin{equation}
\label{RGEqnAnHL}
\frac{d\,\ln \bigl[\tal{(\ell)}{an}(\mu^2)\bigr]}{d\,\ln\mu^2} =
- \left\{
\sum_{j=0}^{\ell-1} B_j \Bigl[\tal{(\ell)}{s}(\mu^2)\Bigr]^{j+1}
\right\}_{\mbox{$\!$\scriptsize an}}, \qquad
B_j = \frac{\beta_j}{\bz^{j+1}},
\end{equation}
where $\al{(\ell)}{s}(\mu^2)$ is the $\ell$-loop perturbative running
coupling, $\tal{}{}(\mu^2)=\al{}{}(\mu^2) \bz/(4 \pi)$, and
$\beta_{j}$ are the $\beta$~function expansion coefficients
(see~\ref{Sect:RCPert}). By employing the relation~(\ref{DefAn}) the
right-hand side of Eq.~(\ref{RGEqnAnHL}) can be represented  as a
spectral integral
\begin{equation}
\label{RGEqnAnHL1}
\frac{d\,\ln \bigl[\tal{(\ell)}{an}(\mu^2)\bigr]}{d\,\ln\mu^2} =
- \int\limits_{0}^{\infty}\frac{\Ro{\ell}}{\sigma+\mu^2}\, d\sigma,
\end{equation}
where
\begin{equation}
\label{SpFunDef}
\Ro{\ell} = \frac{1}{2 \pi i}\,
\lim_{\varepsilon \to 0_{+}} \sum_{j=0}^{\ell-1} B_j
\left\{\left[\tal{(\ell)}{s}(-\sigma-i\varepsilon)\right]^{j+1}
- \left[\tal{(\ell)}{s}(-\sigma+i\varepsilon)\right]^{j+1}\right\}.
\end{equation}
The explicit forms of ${\cal R}^{(\ell)}(\sigma)$ at different loop
levels are given in~\ref{Sect:SF}. Integrating Eq.~(\ref{RGEqnAnHL1})
with respect to~$\mu^2$ in finite terms, we obtain
\begin{equation}
\label{AICHLNorm}
\al{(\ell)}{an}(q^2) = \al{(\ell)}{an}(q_0^2)\,
\exp\!\left[\int\limits_{0}^{\infty}\!\Ro{\ell}
\,\ln\!\left(\frac{1+\sigma/z}{1+\sigma/z_0}\right)
\frac{d \sigma}{\sigma}\right],
\end{equation}
where $z=q^2/\Lambda^2$ and $z_0=q_0^2/\Lambda^2$.

     Expression~(\ref{AICHLNorm}) contains the explicit dependence on
the normalization point~$q_0^2$ and on the value of the strong
running coupling at this point. It is worth mentioning here that this
equation possesses the obvious renormalization invariance. In fact,
having the value of the QCD running coupling at some momenta
transferred, one might exploit Eq.~(\ref{AICHLNorm}). However,
usually it turns out to be more convenient to deal with the explicit
expression for the strong running coupling independent of the
normalization point.

     For the case of the model in hand this objective can be achieved
by invoking the physical requirement\footnote{This condition has
also been used in Ref.~\refcite{France01} for evaluating the
normalization coefficients.} $\al{(\ell)}{an}(q^2) \to
\al{(\ell)}{s}(q^2)$, when $q^2 \to \infty$ (see Refs.~\refcite{PRD2}
and~\refcite{MPLA2} for details). In fact, this condition has
already been employed at  the one-loop level, see
Eqs.~(\ref{AIC1LNorm}) and~(\ref{AIC1L}). It is worth emphasizing
here that the imposition of such requirement does not affect the
renormalization invariance of Eq.~(\ref{AICHLNorm}). In general,
one is able to derive the integral representation for the QCD
analytic invariant charge, independent of the normalization point, by
invoking the similar requirement $\al{(\ell)}{an}(q^2) \to
\al{(1)}{an}(q^2)$, when $q^2 \to \infty$. For this purpose let us
consider the ratio of the $\ell$-loop analytic running
coupling~(\ref{AICHLNorm}) normalized at a point~$q_0^2$ to the
one-loop AIC written in the form~(\ref{AICHLNorm}) and  normalized at
the same point~$q_0^2$:
\begin{equation}
\frac{\al{(\ell)}{an}(q^2)}{\al{(1)}{an}(q^2)} =
\frac{\al{(\ell)}{an}(q_0^2)}{\al{(1)}{an}(q_0^2)}\,
\exp\!\left[
\int\limits_{0}^{\infty}\! \DRo{\ell}{\sigma} \,
\ln\!\left(\frac{1 + \sigma/z}{1 + \sigma/z_0}\right)
\frac{d \sigma}{\sigma}\right],
\end{equation}
where $\DRo{\ell}{\sigma} = \Ro{\ell} - \Ro{1}$. Proceeding to
the limit $q_0^2 \to \infty$, one arrives at the expression for the
$\ell$-loop analytic invariant charge\cite{PRD2,MPLA2}
\begin{equation}
\label{AICHL}
\al{(\ell)}{an}(q^2) = \frac{4\pi}{\beta_0}\,\frac{z-1}{z \, \ln z}\,
\exp\!\left[\int\limits_{0}^{\infty}\! \DRo{\ell}{\sigma}\,
\ln\!\left(1 + \frac{\sigma}{z}\right) \frac{d \sigma}{\sigma}\right],
\qquad z=\frac{q^2}{\Lambda^2}.
\end{equation}

     It is worthwhile to note that the integral representation of the
\KL type also holds for the AIC~(\ref{AICHL})
\begin{equation}
\label{AICHLKL}
\al{(\ell)}{an}(q^2) = \frac{4\pi}{\beta_0} \int\limits_{0}^{\infty}
\frac{\ro{\ell}(\sigma)}{\sigma + z}\, d \sigma,
\end{equation}
where $\ro{\ell}(\sigma)$ is the $\ell$-loop spectral density
\begin{eqnarray}
\ro{\ell}(\sigma) &=& \ro{1}(\sigma) \,
\exp\!\left[ \pvint{0}{\infty} \DRo{\ell}{\zeta} \,
\ln \left| 1 - \frac{\zeta}{\sigma}\right| \,
\frac{d \zeta}{\zeta} \right] \nonumber \nopagebreak \\
&& \times \left[\cos \psi^{(\ell)}(\sigma) +
\frac{\ln \sigma}{\pi} \sin \psi^{(\ell)}(\sigma) \right],
\label{SpDnsHL}
\end{eqnarray}
the one-loop spectral density $\ro{1}(\sigma)$ has been specified in
Eq.~(\ref{SpDns1L}), and
\begin{equation}
\label{PsiDef}
\psi^{(\ell)}(\sigma) = \pi \int\limits_{\sigma}^{\infty}\!
\DRo{\ell}{\zeta} \, \frac{d \zeta}{\zeta}.
\end{equation}
In the exponent of Eq.~(\ref{SpDnsHL}) the principle value of  the
integral is assumed. However, in practical use the
expression~(\ref{AICHL}) turns out to be more convenient. One  should
emphasize here that, similarly to the one-loop case considered above,
the analytic invariant charge~(\ref{AICHL}) has no unphysical
singularities, it contains no free parameters, and  it incorporates
the ultraviolet asymptotic freedom with the infrared enhancement in
an unified manner.

\begin{figure}
\centerline{\psfig{file=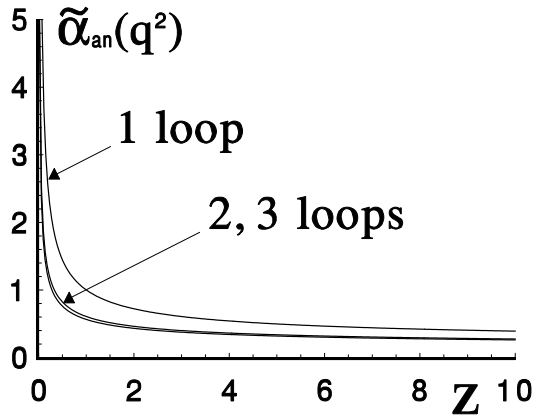,width=70mm}}
\vspace*{8pt}
\caption{The QCD analytic invariant charge~(\protect\ref{AICHL}) at
different loop levels, $z = q^2 / \Lambda^2$.}
\label{Plot:AICHL}
\end{figure}

     Figure~\ref{Plot:AICHL} shows the analytic invariant
charge~(\ref{AICHL}) $\tal{(\ell)}{an}(q^2) = \al{(\ell)}{an}(q^2)\,
\bz/(4 \pi)$ at the one-, two-, and three-loop levels. It is obvious
that AIC possesses the higher loop stability. Thus, the curves
corresponding to the two- and three-loop levels are practically
indistinguishable. Proceeding from this one can also draw the
conclusion concerning the scheme stability of the current approach.
In particular, in Fig.~\ref{Plot:AICHL} the curve corresponding to
the three-loop analytic running coupling $\tal{(3)}{an}(q^2)$ is
plotted by making use of the coefficient $\beta_2 = 2857/2 - 5033\,
\nf/18 + 325\, \nf^2/54$ computed in the \MSbar scheme.\cite{TVZ} The
account of the third term on the right-hand side of
Eq.~(\ref{AnRGEq}) does not lead to a valuable quantitative variation
of its solution in comparison with the two-loop approximation.
Therefore it is evident that using the coefficient $\beta_2$ computed
in another subtraction scheme\footnote{At least, in schemes that do
not have unnaturally large expansion coefficients (see
paper~\refcite{Rac} and references therein for the detailed
discussion of this issue).} does not lead to significant variation of
the solution to Eq.~(\ref{AnRGEq}) in comparison with the considered
case of the \MSbar scheme. This statement follows also from the fact
that the contribution of every subsequent term on the right-hand side
of Eq.~(\ref{AnRGEq}) is substantially suppressed by the
contributions of the preceding ones (see also Ref.~\refcite{MPLA2}
and~\ref{Sect:SF}).

\subsection{Properties of the AIC at higher loop levels}
\label{Sect:AICHLProp}

     At the one-loop level there is an explicit expression for the
analytic invariant charge~(\ref{AIC1L}). This allows one to perform
the investigation of its properties manifestly. However, at higher
loop levels only the integral representation~(\ref{AICHL}) has been
obtained for $\al{(\ell)}{an}(q^2)$ so far. This fact significantly
complicates the issue in hand, and eventually leads to  the necessity
of applying the numerical computation. Nevertheless, the study of the
relevant $\beta$~function enables one to elucidate some important
questions here, in particular, the asymptotic behavior of the
analytic invariant charge (see Ref.~\refcite{MPLA2} for details).

\begin{figure}[t]
\centerline{\psfig{file=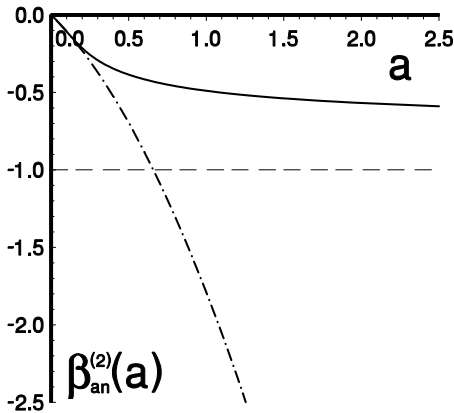,width=60mm}}
\vspace*{8pt}
\caption{The $\beta$~function corresponding to the two-loop analytic
invariant charge (solid curve). The relevant perturbative result is
shown as the dot-dashed curve.}
\label{Plot:Beta2L}
\end{figure}

\begin{figure}[t]
\centerline{\psfig{file=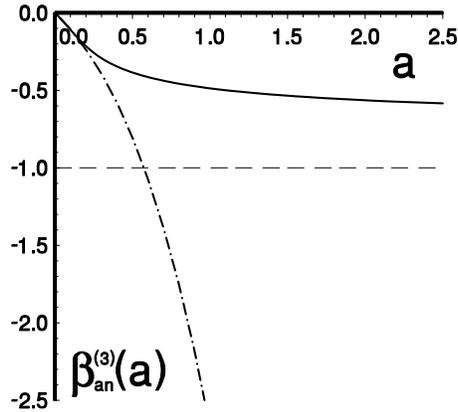,width=60mm}}
\vspace*{8pt}
\caption{The $\beta$~function corresponding to the three-loop analytic
invariant charge (solid curve). The relevant perturbative result is
shown as the dot-dashed curve.}
\label{Plot:Beta3L}
\end{figure}

     The $\beta$~functions corresponding to the analytic invariant
charge~(\ref{AICHL}) at the higher loop levels
\begin{equation}
\label{BetaHLDef}
\bt{(\ell)}{an}(a) = \frac{d \, \ln a(\mu^2)}{d \, \ln \mu^2}, \qquad
a(\mu^2) = \tal{(\ell)}{an}(\mu^2)
\end{equation}
are presented in Figures~\ref{Plot:Beta2L} and~\ref{Plot:Beta3L}.
Figure~\ref{Plot:Beta2L} shows the two-loop  $\beta$~function
$\bt{(2)}{an}(a)$ together with respective perturbative result
$\bt{(2)}{s}(a)=-a-B_1 a^2$. Figure~\ref{Plot:Beta3L} depicts the
analogous functions at the three-loop level, namely $\bt{(3)}{an}(a)$
and $\bt{(3)}{s}(a)=-a-B_1 a^2-B_2 a^3$. It is clear from Figures
\ref{Plot:Beta1L}, \ref{Plot:Beta2L}, and~\ref{Plot:Beta3L} that the
$\beta$~function corresponding to the analytic running
coupling~(\ref{AICHL}) coincides with its perturbative analog in the
region of small values of the invariant charge at any loop level. In
other words, in the framework of the model under consideration the
complete recovering of the perturbative limit in the ultraviolet
domain takes place.

\begin{figure}[t]
\centerline{\psfig{file=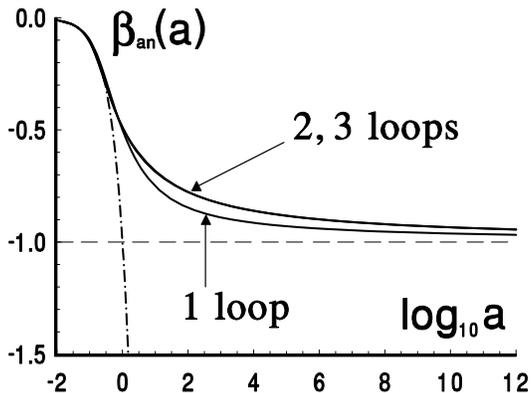,width=70mm}}
\vspace*{8pt}
\caption{The $\beta$ function corresponding to the analytic invariant
charge~(\protect\ref{AICHL}) at different loop levels (solid curves).
The one-loop perturbative result is shown as the dot-dashed curve.}
\label{Plot:BetaHL}
\end{figure}

     Let us turn now to the asymptotics of the $\beta$~function
(\ref{BetaHLDef}) at large values of the running coupling.
Figure~\ref{Plot:BetaHL} presents the $\beta$ functions
$\bt{(\ell)}{an}(a),\,$ $\ell = 1,2,3$ and the one-loop perturbative
result $\bt{(1)}{s}(a) = -a$. This figure clearly shows the
perturbative limit at small $a$, as well as the universal asymptotic
behavior of the $\beta$ function in hand: $\bt{(\ell)}{an}(a) \to
-1$, when $a \to \infty$. The latter statement can also be proved in
an independent way. Indeed, due to the infrared enhancement of the
analytic running coupling, the value of the right-hand side of
Eq.~(\ref{RGEqnAnHL}) (it is nothing but the $\beta$ function at the
relevant loop level) when $\mu^2 \to 0$ corresponds to the limit
$\tal{(\ell)}{an}(\mu^2) \to \infty$. One can show that irrespective
of the loop level
\begin{equation}
\lim_{\,\mu^2 \to 0}
\left\{\tal{(\ell)}{s}(\mu^2)\right\}_{\!\mbox{\scriptsize an}} = 1,
\qquad
\lim_{\,\mu^2 \to 0} \left\{\left[
\tal{(\ell)}{s}(\mu^2)\right]^{j+1}\right\}_{\!\mbox{\scriptsize an}} = 0,
\end{equation}
where $j$ is a natural number ($j=1, 2, 3, ...$). Therefore, we infer
that at any loop level the $\beta$~function~(\ref{BetaHLDef})
corresponding to the analytic invariant charge~(\ref{AICHL}) tends to
the universal limit
\begin{equation}
\lim_{a \to \infty}\, \bt{(\ell)}{an}(a) = -1.
\end{equation}
As it has been mentioned above, such a behavior of the $\beta$
function leads to the IR enhancement of the running coupling, namely
$\tal{}{}(q^2) \simeq \Lambda^2/q^2$, when $q^2 \to 0$.

     Thus, the $\beta$ function corresponding to the analytic
invariant charge~(\ref{AICHL}) has the universal asymptotic behaviors
at both small ($\bt{(\ell)}{an}(a) \simeq -a$, when $a \to 0$) and
large  ($\bt{(\ell)}{an}(a)\simeq -1$, when $a \to \infty$)  values
of the running coupling irrespective of the loop level. Therefore,
the analytic invariant charge itself possesses the universal
asymptotics both in the ultraviolet ($\tal{(\ell)}{an}(q^2) \simeq
1/\ln (q^2/\Lambda^2)$, when $q^2\to\infty$) and infrared
($\tal{(\ell)}{an}(q^2) \simeq \Lambda^2/q^2$, when $q^2 \to 0$)
regions at any loop level. In particular, this implies that using the
AIC~(\ref{AICHL}) at the higher loop levels will also result in the
confining quark--antiquark potential (see Subsection~\ref{Sect:Vr}).

\subsection{Extension to the timelike region}
\label{Sect:TimeLike}

     In the previous sections the model for the QCD analytic
invariant charge has been studied in the spacelike (Euclidean)
region. However, for congruous description of a number of strong
interaction processes (for example, inclusive $\tau$~lepton decay or
electron-positron annihilation into hadrons) one has to use the
continuation of the invariant charge to the timelike (Minkowskian)
region. In general, one might expect such a continuation to affect
the properties of the strong running coupling. Therefore, it is of a
certain importance to extend the analytic invariant charge to the
timelike domain, and to compare the functions obtained.

     The consistent description of hadron dynamics in spacelike and
timelike domains remains the goal of many studies for quite a long
time. Thus, the first attempts to build up the expression for the
strong running  coupling valid for the timelike momenta transferred
were made in the early 1980s (see Refs.~\refcite{Schrempp80}--\refcite{PRKP}).
Recently, in Ref.~\refcite{APTTL} the procedure
of continuation of the invariant charge from the spacelike to the
timelike region (and vice versa) was elaborated by making use of the
dispersion relation for the Adler~$D$ function. In the framework of
this method the integral relationships between the QCD running
coupling in Euclidean and  Minkowskian domains were established. In
particular, the continuation  of the strong running coupling to the
timelike region is connected with the ``spacelike'' invariant charge
through the relation (see paper~\refcite{APTTL} and references therein)
\begin{equation}
\label{RCTLDef}
\hal{}{}(s) = \frac{1}{2 \pi i}
\int\limits_{s+i\varepsilon}^{s-i\varepsilon}
\alpha(-\zeta)\,\frac{d \zeta}{\zeta}, \qquad s=-q^2>0.
\end{equation}
In this equation the integration contour goes from the point
$s+i\varepsilon$ to the point $s-i\varepsilon$ and lies in the region
of analyticity of the function $\alpha(-\zeta)$.  Here and further
the running coupling in the spacelike region is denoted by
$\alpha(q^2)$, and in the timelike region by $\hal{}{}(s)$. It is
worth mentioning also that the relation inverse to~(\ref{RCTLDef}) is
\begin{equation}
\alpha(q^2) = q^2 \int\limits_{0}^{\infty}
\frac{\widehat{\alpha}(s)}{(s + q^2)^2} \, d s.
\end{equation}

\begin{figure}
\centerline{\psfig{file=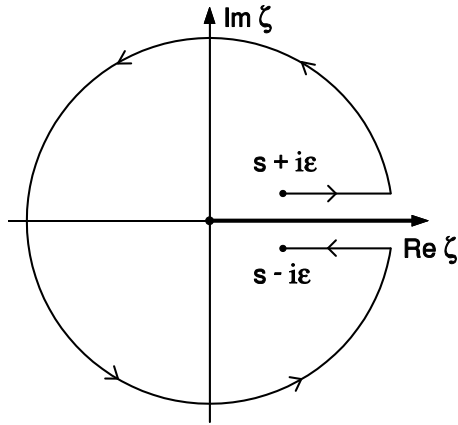,width=60mm}}
\vspace*{8pt}
\caption{The integration contour in Eq.~(\protect\ref{RCTLDef}).
The physical cut of the invariant charge $\alpha(-\zeta)$ is shown
along the positive semiaxis of real~$\zeta$.}
\label{Plot:ContTL}
\end{figure}

     It is convenient to choose the integration contour in
Eq.~(\ref{RCTLDef}) in the form presented in
Figure~\ref{Plot:ContTL}. Namely, the integration path starts from
the point $s+i\varepsilon$, proceeds in a parallel way with the real
axis to infinity, then along the circle of infinitely large radius it
goes counter-clockwise to the point $(\infty-i\varepsilon)$, and then
it goes in a parallel way with the real axis to the point
$s-i\varepsilon$. Since the QCD running coupling possesses the
property of asymptotic freedom (namely, $\tal{}{}(q^2)\simeq
1/\ln(q^2/\Lambda^2)$, when $q^2\to\infty$), the integral along the
circle of an infinitely  large radius gives vanishing contribution to
the right-hand side of Eq.~(\ref{RCTLDef}). Then, the remaining
expression can be reduced to\cite{APTTL}
\begin{equation}
\widehat{\alpha}(s) = \int\limits_{s}^{\infty}\varrho(\zeta)
\,\frac{d \zeta}{\zeta},
\end{equation}
where $\varrho(\zeta)$ denotes the spectral density of the \KL
representation for the running coupling~$\alpha(q^2)$:
\begin{equation}
\varrho(\zeta) = \frac{1}{2\pi i}\,
\lim_{\varepsilon \to 0_{+}}
\Bigl[\alpha(-\zeta-i\varepsilon) -
\alpha(-\zeta+i\varepsilon)\Bigr].
\end{equation}
Thus, the continuation of the analytic invariant charge~(\ref{AICHL})
to the timelike region is given by
\begin{equation}
\label{AICTL}
\hal{(\ell)}{an}(s) = \frac{4\pi}{\beta_{0}}\,
\int\limits_{s/\Lambda^2}^{\infty}~\!\!
\ro{\ell}(\zeta)\,\frac{d \zeta}{\zeta}
\end{equation}
with the spectral density $\ro{\ell}(\zeta)$ denoted in
Eqs.~(\ref{SpDnsHL}) and (\ref{PsiDef}) (see Ref.~\refcite{PRD2} for
details).

\begin{figure}[t]
\centerline{\psfig{file=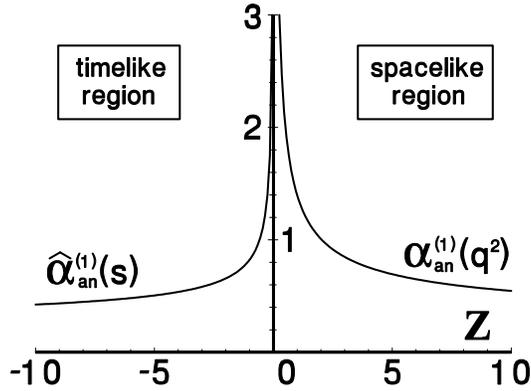,width=70mm}}
\vspace*{8pt}
\caption{The one-loop analytic running coupling in the spacelike
($q^2>0$) and timelike ($s=-q^2>0$) regions, $z=q^2/\Lambda^2$.}
\label{Plot:AICTL}
\end{figure}

     The one-loop running coupling~(\ref{AICTL}) has the following
asymptotic in the ultraviolet limit:
\begin{equation}
\label{AICTLUV}
\hal{(1)}{an}(s) \simeq \frac{4\pi}{\beta_{0}} \, \frac{1}{\ln z} \,
\left\{1 - \frac{\pi^2}{3} \frac{1}{(\ln z)^2} +
{\cal O}\!\left[\frac{1}{(\ln z)^4},
\frac{1}{z}\right]\right\}, \qquad s \to \infty,
\end{equation}
where $z=s/\Lambda^2$. On the one hand, this running coupling has
correct UV behavior determined by the asymptotic freedom. On the
other hand, the so-called $\pi^2$-terms have also appeared in
expansion~(\ref{AICTLUV}). These terms play a key role in  the
description of the strong interaction processes in the timelike
domain (see discussion of this issue in
Refs.~\refcite{APTRev,APTTL,Pi1}, and~\refcite{Pi2}). It is
worthwhile to mention also that, similarly to the ``spacelike'' case,
there is the IR enhancement of the one-loop running
coupling~(\ref{AICTL}):
\begin{equation}
\label{AICTLIR}
\hal{(1)}{an}(s) \simeq \frac{4\pi}{\beta_{0}}\,\frac{1}{z\,(\ln z)^2},
\qquad s \to 0.
\end{equation}
However, here the type of the infrared singularity differs from that
of the invariant charge~(\ref{AIC1L}). Nevertheless, it is this
behavior of the running coupling~(\ref{AICTLIR}) that enables one to
handle the integrals over the infrared domain of the form (see, e.g.,
Ref.~\refcite{DWKT})
\begin{equation}
\bar{A}(s) = \frac{1}{s}\int\limits_{0}^{s}\hal{}{an}(s')\, d s'.
\end{equation}

     It is interesting to note also that within the current approach
a simple relation holds between the $\beta$~function corresponding to
the running coupling in the timelike region~(\ref{AICTL}) and the
relevant spectral density:
\begin{equation}
\beta\Bigl(\hal{}{}(s)\Bigr) =
\frac{d\, \bigl[\hal{(\ell)}{an}(s)\bigr]}{d\,\ln s} =
- \frac{4\pi}{\beta_{0}} \, \ro{\ell}(s).
\end{equation}
Thus, the $\beta$~function in hand proves to be proportional to the
spectral density~$\ro{\ell}(s)$. Obviously, this result justifies the
attempts, originated by Schwinger,\cite{Schwinger} to find a direct
physical interpretation of the $\beta$ function (the so-called
Schwinger's hypothesis, see also Refs.~\refcite{APTTL}
and~\refcite{Milton}).

\begin{figure}[t]
\centerline{\psfig{file=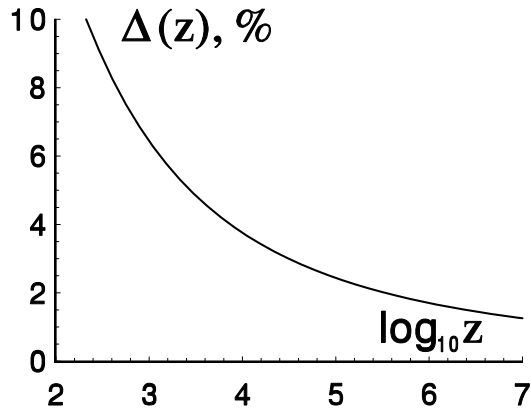,width=70mm}}
\vspace*{8pt}
\caption{The relative difference $\Delta (z) = \bigl[
\al{(1)}{an}(q^2)/\hal{(1)}{an}(-q^2)-1\bigr]\times 100\,\%$ between
the values of the one-loop analytic running coupling in  the
spacelike and timelike regions, $z=q^2/\Lambda^2$.}
\label{Plot:STDiff}
\end{figure}

     The plots of the functions $\al{(1)}{an}(q^2)$ and
$\hal{(1)}{an}(s)$ are shown in Figure~\ref{Plot:AICTL}. In the
ultraviolet limit these  expressions have identical behavior
determined by the asymptotic freedom.  However, there is asymmetry
between them in the intermediate- and low-energy regions. The ratio
of the one-loop analytic running coupling in the spacelike
region~$\al{(1)}{an}(q^2)$ to its continuation to the timelike region
$\hal{(1)}{an}(-q^2)$ is presented in Figure~\ref{Plot:STDiff}.
The relative difference between these functions is about several
percent at the scale of the $Z$~boson mass, and increases when
approaching the infrared domain. Apparently, this circumstance must
be taken into account when one handles the experimental data. In
particular, this will play a crucial role in description of the
inclusive $\tau$~lepton decay and the  $e^{+}e^{-}$~annihilation
into hadrons in the framework of the approach  developed (see
Subsections \ref{Sect:Tau} and \ref{Sect:EpEm}, respectively).

% SECTION 4

\section{Phenomenological Applications}
\label{Sect:Applic}

     As has been mentioned above, there are several ways of
incorporating the analyticity requirement into the renormalization
group method. Eventually, this leads to different models for the
running coupling in the framework of the analytic approach to QCD.
Certainly, every such model, as well as any new model for the strong
interaction, must be verified thoroughly. A decisive test of a
model's self--consistency is its applicability to description of
diverse QCD processes. Since we are working within a nonperturbative
approach, the study of intrinsically nonperturbative phenomena is of
a primary importance here. Furthermore, the elaborated model for the
analytic invariant charge contains no free parameters, i.e.,
similarly to perturbative approach, \LQCD remains the basic
characterizing parameter of the theory. Hence, the congruity of the
estimations of this parameter, obtained for different strong
interaction processes, would imply the quantitative consistency of
the developed model (see also papers~\refcite{France01,ConfV,LCEPP01}
and references therein).

\subsection{Static quark--antiquark potential}
\label{Sect:Vr}

     Let us proceed to construction of the static quark--antiquark
potential in the framework of the one-gluon exchange model. This
potential is related to the strong running coupling $\alpha(q^2)$ by
the three-dimensional Fourier transformation
\begin{equation}
\label{VrGen}
V(r) = - \frac{16 \pi}{3} \int\limits_{0}^{\infty}
\frac{\alpha(\bm{q}^2)}{\bm{q}^2}\,
\frac{\exp(i \bm{q r})}{(2 \pi)^3}\, d {\bm q}
\end{equation}
(see, e.g., reviews~\refcite{Nora,Kiselev} and references
therein for details). In fact, the lowest--lying bound states of
heavy quark systems may well be described by the
potential~(\ref{VrGen}) even with the perturbative expression for the
QCD running coupling.\cite{Ynd1} However, at large distances, which
play the crucial role in the hadron spectroscopy, the perturbative
approach blows up due to unphysical singularities (such as the Landau
pole) of the strong  running coupling.

     For the construction of the quark--antiquark
($\mbox{Q}\overline{\mbox{Q}}$) potential we shall use here the
analytic invariant charge (\ref{AIC1L}) (see also
Refs.~\refcite{Potent,PRD1,ConfIV}, and~\refcite{LCEPP99}).  After
integration over the angular variables, Eq.~(\ref{VrGen}) acquires
the form
\begin{equation}
\label{VrInt}
V(r) = - \frac{32}{3 \beta_0}\, \frac{1}{r_0} \int\limits_{0}^{\infty}
\frac{p^2-1}{p^2 \ln p^2}\, \frac{\sin (p R)}{p R}\, d p,
\end{equation}
where $p = q r_0$, $R = r/r_0$, and $r_0$ is a reference scale of the
dimension of length, which will be specified below. This equation is
rather complicated to be integrated explicitly. Nevertheless, one is
able to study the asymptotics of the  quark--antiquark
potential~(\ref{VrInt}) by making use of the following
prescription. First of all, let us introduce a dimensionless
variable $Q = p R$ and rewrite Eq.~(\ref{VrInt}) as follows:
\begin{equation}
\label{VrInt1}
V(r) = \frac{16}{3 \beta_0}\, \frac{1}{r_0} \, \frac{1}{R}
\int\limits_{0}^{\infty}\!\left(1 - \frac{R^2}{Q^2}\right)
\frac{\sin Q}{Q}\, \frac{d Q}{\ln R-\ln Q}.
\end{equation}
Then, formally expanding the denominator of the integrand,\footnote{A
similar method has also been used in Ref.~\refcite{LeTo}.} one can
find the asymptotic behavior of $V(r)$ at both small and large
distances:
\begin{eqnarray}
V(r) &\simeq& \frac{16}{3 \beta_0}\,\frac{1}{r_0}
\Biggl[
\frac{1}{R} \sum_{n=0}^{n_0}\frac{1}{(\ln R)^{n+1}}
\int\limits_{0}^{\infty} \frac{\sin Q}{Q} (\ln Q)^n\, d Q
\nonumber \nopagebreak \\ \nopagebreak
& & - R \sum_{m=0}^{m_0}\frac{1}{(\ln R)^{m+1}}
\int\limits_{0}^{\infty} \frac{\sin Q}{Q^3} (\ln Q)^m\, d Q
\Biggr].
\label{VrIntExp}
\end{eqnarray}
The values of $n_0$ and $m_0$ will be specified below.

     For evaluation of the expansion coefficients in
Eq.~(\ref{VrIntExp}) it is worth considering an auxiliary integral of
the form
\begin{equation}
\label{IntAux1}
\int\limits_{0}^{\infty} Q^t\, \sin Q \, d Q = \sqrt{\pi}\, 2^t \,
\frac{\Gamma\left(1+\frac{t}{2}\right)}
{\Gamma\left(\frac{1}{2} - \frac{t}{2}\right)}.
\end{equation}
Differentiating this equation $n$ times with respect to variable $t$
we obtain
\begin{equation}
\label{IntAux2}
\int\limits_{0}^{\infty} Q^t\, (\ln Q)^n \, \sin Q \, d Q = v(n, t),
\end{equation}
where
\begin{equation}
\label{vDef}
v(n, t) = \sqrt{\pi}\, \frac{d^n}{d t^n} \left[2^t\,
\frac{\Gamma\left(1+\frac{t}{2}\right)}
{\Gamma\left(\frac{1}{2} - \frac{t}{2}\right)}
\right].
\end{equation}
In Eq.~(\ref{IntAux1}) the parameter $t$ takes the values
$0\le|\Re(t+1)|<1$ (see, e.g., Ref.~\refcite{RG}). However, in order
to determine the expansion coefficients of the second sum in
Eq.~(\ref{VrIntExp}), one has to go to the point $t=-3$, which is
outside of this range. Nevertheless, it can be done by making use of
the analytic continuation of the left-hand side of
Eq.~(\ref{IntAux1}). This continuation is unique and it is defined
obviously by the right-hand side of Eq.~(\ref{IntAux1}) all over  the
complex $t$--plane except for the points $t=-2 N$, with  $N$ being a
natural number. Fortunately, we are not dealing with these values of
the parameter~$t$, and we can put
\begin{equation}
\label{IntCont}
\int\limits_{0}^{\infty} \left.\frac{\sin Q}{Q^s} (\ln Q)^n\, d Q =
v(n, t)\right|_{t=-s}, \quad s \neq 2 N\quad (N=1,2,3, ... ).
\end{equation}
It should be noted here that this analytical continuation plays the
role of regularization of the expansion coefficients in
Eq.~(\ref{VrIntExp}). It is also convenient to introduce the
notations
\begin{equation}
\label{uwDef}
u_{n} = \left.\frac{2}{\pi}\, v(n, t)\right|_{t=-1}, \qquad
\omega_{m} = -\left.\frac{2}{\pi}\, v(m, t)\right|_{t=-3}.
\end{equation}
The explicit expressions for these coefficients can be found
in~\ref{Sect:VrExp}.

     Thus, the static quark--antiquark potential (\ref{VrIntExp})
can be represented now in the following form:
\begin{eqnarray}
\label{VrAsympt}
V(r) \simeq \frac{8 \pi}{3 \beta_0}\, \frac{1}{r_0} \left[
\frac{1}{R} \sum_{n=0}^{n_0} \frac{u_n}{(\ln R)^{n+1}} +
R \sum_{m=0}^{m_0} \frac{\omega_m}{(\ln R)^{m+1}}\right],
\qquad R=\frac{r}{r_0}.
\end{eqnarray}
At small distances the potential~(\ref{VrAsympt}) possesses the
standard behavior, determined by the asymptotic freedom
\begin{equation}
V(r) = \frac{8 \pi}{3 \beta_0}\, \frac{1}{r_0}\,
\frac{1}{R\, \ln R}, \qquad r \to 0.
\end{equation}
At the same time, it proves to be rising at large distances
\begin{equation}
V(r) = \frac{8 \pi}{3 \beta_0}\, \frac{1}{r_0}\,
\frac{R}{2\, \ln R}, \qquad r \to \infty,
\end{equation}
implying the confinement of quarks. It is of a particular interest
to mention here that a similar rising behavior of the~\QbQ potential
has been proposed\cite{FLW} a long time ago proceeding from the
phenomenological assumptions.

     As it has been shown in Subsection~\ref{Sect:AICHLProp}, the
analytic invariant charge~(\ref{AICHL}) has the universal asymptotics
both in the ultraviolet and infrared regions at any loop level.
Furthermore, the approach in hand possesses a good scheme stability.
Therefore, neither higher loop corrections, nor scheme dependence,
can affect qualitatively the results obtained. The use of different
subtraction schemes (or different loop approximations) will result in
redefinition of the scale parameter, as it also takes place within
the perturbative approach.

     Equation (\ref{VrAsympt}) describes the behavior of the static
quark--antiquark potential~(\ref{VrInt}) at small and large
distances. However, its straightforward extrapolation to all
distances encounters poles of different orders at the point~$R=1$,
which apparently is an artifact of the expansion~(\ref{VrIntExp}).
For the practical purposes it would be undoubtedly useful to derive
an explicit expression for the \QbQ potential applicable for $0 < r <
\infty$.

     In order to construct an explicit interpolating formula for the
quark--antiquark potential~(\ref{VrInt}) we shall employ  here the
following method. Let us modify the expansion (\ref{VrAsympt}) in a
``minimal'' way, by adding terms which only subtract the
singularities at the point $R=1$ and do not contribute to the derived
asymptotics. This leads to the following expression for the static
quark--antiquark potential
\begin{equation}
\label{VrReg}
V(r) = \frac{8 \pi}{3 \beta_0}\, \frac{1}{r_0} \left\{
\sum_{n=0}^{n_0} u_n
\left[\frac{1}{R\, (\ln R)^{n+1}} \right]_\Reg +
\sum_{m=0}^{m_0} \omega_m
\left[ \frac{R}{(\ln R)^{m+1}}\right]_\Reg\right\},
\end{equation}
where $R=r/r_0$. The functions $[1/(R\,(\ln R)^n)]_\Reg$ and
$[R/(\ln R)^m]_\Reg$ are presented in an explicit form
in~\ref{Sect:VrExp}.

\begin{figure}
\centerline{\psfig{file=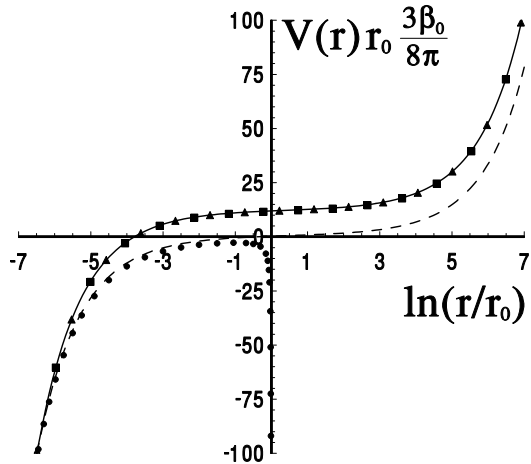,width=70mm}}
\vspace*{5.5pt}
\caption{The quark--antiquark potential (\protect\ref{VrReg})  in
dimensionless units at different levels of approximation: $n_0 = m_0
= 0$ (dashed curve), $n_0 = m_0 = 4$ (solid curve), $n_0 = m_0 = 5$
($\blacktriangle$), and $n_0 = m_0 = 10$ ({\tiny $\blacksquare$}).
The one-loop perturbative result is shown by dotted curve.}
\label{Plot:VrAppr}
\end{figure}

     The numerical analysis of Eq.~(\ref{VrReg}) revealed that for
practical purposes it is enough to retain the first five expansion
terms ($n_0 = m_0 = 4$) therein. It turns out that the
potential~(\ref{VrReg}) itself and the  corresponding estimation of
the value of parameter \LQCD are not affected by higher--order
contributions. In particular, the curves (\ref{VrReg}) for $n_0 = m_0
= 5$ and for  $n_0 = m_0 = 10$ are practically indistinguishable of
the curve corresponding to $n_0 = m_0 = 4$ over the whole region
$0<r<\infty$ (see Figure~\ref{Plot:VrAppr}). And the higher--order
estimations of the parameter \LQCD vary within $0.5\,\%$ of the value
obtained at the $n_0 = m_0 = 4$.

     Thus, we arrive at the following explicit expression for the
static quark--antiquark potential:\cite{Potent}
\begin{eqnarray}
V(r) &=& V_{0} + \frac{8 \pi}{3 \beta_0}\, \frac{1}{r_0} \Biggl\{
\frac{1}{R}\Biggl[
\frac{1}{\ln R} - \frac{0.577}{(\ln R)^2}
+ \frac{1.156}{(\ln R)^3}
- \frac{4.021}{(\ln R)^4} + \frac{15.018}{(\ln R)^5}\Biggr]
\nonumber \\ &&
+ R \Biggl[\frac{0.500}{\ln R} + \frac{0.461}{(\ln R)^2}
+ \frac{1.462}{(\ln R)^3} + \frac{3.185}{(\ln R)^4}
+ \frac{17.844}{(\ln R)^5}\Biggr]
\nonumber \\ &&
+ \frac{21.489}{1-R}
- \frac{62.484}{(1-R)^2}
+ \frac{98.694}{(1-R)^3} - \frac{84.143}{(1-R)^4}
+ \frac{32.861}{(1-R)^5} \Biggr\},
\label{Vr}
\end{eqnarray}
where $R=r/r_0$. In order to reproduce the correct short-distance
behavior of the \QbQ potential, the dimensional parameter~$r_0$ in
this equation has to be identified with
$\Lambda_{\overline{\mbox{\tiny MS}}}$ by the relation
$r_0^{-1} = \Lambda \exp(\gamma)$,  where $\gamma \simeq 0.57721...$
denotes the Euler's constant (see, e.g.,  Refs.~\refcite{Peter,KKO}
and~\refcite{Melles}).  It is straightforward to verify that the
potential (\ref{Vr}) satisfies also the concavity condition
\begin{equation}
\frac{d\,V(r)}{d\,r} > 0, \qquad \frac{d^2 V(r)}{d\, r^2} \le 0,
\end{equation}
which is a general property of the gauge theories (see
Ref.~\refcite{SB} for details).

\begin{figure}
\centerline{\psfig{file=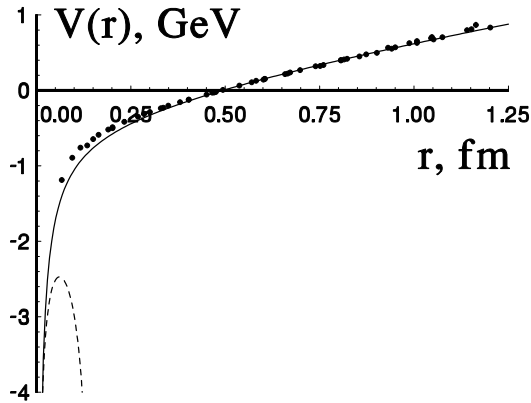,width=70mm}}
\vspace*{8pt}
\caption{Comparison of the quark--antiquark potential $V(r)$ defined
by Eq.~(\protect\ref{Vr}) (solid curve) with the quenched lattice
simulation data\protect\cite{Bali4} (\textbullet). The values of
the parameters are: $\Lambda = 670\,$MeV, $\nfs = 0$,
$V_{0}=-3.164\,\mbox{GeV}$. The dashed curve corresponds to the
relevant one-loop perturbative result.}
\label{Plot:Vr}
\end{figure}

     Comparison of the quark--antiquark potential (\ref{Vr}) with the
quenched lattice simulation data\cite{Bali4} has been performed with
the use of the least square method.\cite{Potent} The varied
parameters in Eq.~(\ref{Vr}) were $r_0$ and the additive self--energy
constant~$V_0$. The result of this fit is presented in
Figure~\ref{Plot:Vr}. This figure shows that in the nonperturbative
physically-relevant range $0.3\,\mbox{fm} \lesssim r \lesssim
1.2\,\mbox{fm}$, in which the average quark separations
$\sqrt{\langle r^2\rangle}$ for quarkonia sits,\cite{Bali2}
the \QbQ potential (\ref{Vr}) reproduces the lattice data\cite{Bali4}
fairly well. At the same time, in the region $r \lesssim
0.05\,\mbox{fm}$ the derived potential coincides with the
perturbative result (dashed curve in Fig.~\ref{Plot:Vr}). The
difference between the lattice data and the expression (\ref{Vr}) in
the intermediate range $0.05\,\mbox{fm} \lesssim r \lesssim
0.3\,\mbox{fm}$ can be explained by the presence of additional
nonperturbative contributions at small distances (see, e.g.,
Ref.~\refcite{BaliSS}).

     Thus, the estimation of parameter \LQCD in the course of this
comparison gives $\Lambda^{\!(\nfs=0)} = (670 \pm 8)\,\mbox{MeV}$
(this value corresponds to the one-loop level with $\nf=0$ active
quarks). It is worth noting that the fit with the use of the maximum
likelihood method also gives a similar value~$\Lambda^{(n_f=0)}
\simeq 650\,$MeV. Further, in order to collate the obtained result
with forthcoming estimations, one has to continue it to the region of
three active quarks. It was performed by employing the matching
procedure, and gives\footnote{It is interesting to note here that
similar values of the parameter \LQCD have also been obtained within
different approaches to this issue (see
Refs.~\refcite{FLW}--\refcite{BGT}).} the value\cite{Potent}
$\Lambda^{\!(\nfs=3)} = (590 \pm 10)\,\mbox{MeV}$.

\subsection{Gluon condensate}

     Let us proceed to another nonperturbative aspect of the strong
interaction, namely, to the gluon condensate. The latter is
determined as the matrix element
\begin{equation}
\label{GCDef}
K = \lim_{x \to y}\left< \mbox{vac} \left| \frac{\alpha}{\pi}
:G_{\mu\nu}^a(x) G_{\mu\nu}^a(y): \right| \mbox{vac} \right>,
\end{equation}
the averaging being performed over the ``true physical''  vacuum
state~$\left|\mbox{vac}\right>$ (see Ref.~\refcite{Zakh99} for a recent
review of this issue). In the framework of perturbative approach the
quantity~(\ref{GCDef}) identically equals to zero.\cite{SVZ}
However, lattice simulations\cite{GCLatt} and comparison of the QCD
sum rules with the experimental data on the low--energy hadron
dynamics\cite{SVZ,NarSR,NSVVZ} testify to the nonzero value of gluon
condensate. The quantity~(\ref{GCDef}) is the subject of thorough
studies for a long time (see, e.g., reviews~\refcite{Zakh99,Arb88} and
Refs.~\refcite{AlekArbu,Alek98}). In particular, the integral relation
between the gluon condensate~(\ref{GCDef}) and the nonperturbative
part of the strong running coupling was established in
Refs.~\refcite{RRY,Alek98} and~\refcite{Arb88}:
\begin{equation}
\label{GCDef1}
K = \frac{3}{\pi^3}\int\limits_{0}^{\infty}
\al{}{np}(k^2) k^2\, dk^2,
\end{equation}
where $\al{}{np}(k^2) = \alpha(k^2) - \al{}{s}(k^2)$. Here
$\alpha(k^2)$ is the QCD invariant charge, and $\al{}{s}(k^2)$
denotes the perturbative running coupling.

     Thus, in the framework of the current approach, the gluon
condensate~(\ref{GCDef1}) at the one-loop level takes the form
(see Eq.~(\ref{AIC1L})):
\begin{equation}
\label{GCAIC}
K = - \frac{12}{\beta_0 \pi^2} \Lambda^4
\int\limits_{0}^{\infty} \frac{dz}{\ln z},
\end{equation}
where $z=k^2/\Lambda^2$. However, the integral on the right-hand side
of this equation diverges. In order to regularize it we employ here
the method proposed in Ref.~\refcite{Arb88}. Namely, let us introduce
the cutoff parameter $k_0$, which has the physical meaning  of the
characteristic scale of the nonperturbative effects. The
phenomenological estimation of this parameter gives\cite{Arb88} $k_0
= (700 \pm 100)\,$MeV. Then, assuming the integration in
Eq.~(\ref{GCAIC}) as the principle value one, we obtain
\begin{equation}
\label{GCReg}
K = - \frac{12}{\beta_0 \pi^2} \Lambda^4\,
\mbox{Li}\!\left(\frac{k_0^2}{\Lambda^2}\right),
\end{equation}
where $\mbox{Li}(x)$ denotes the logarithm--integral\cite{RG}
\begin{equation}
\mbox{Li}(x) = \pvint{0}{x}\frac{d t}{\ln t}, \qquad
\Re x > 1.
\end{equation}

     The solution to Eq.~(\ref{GCReg}) with the phenomenological
value of the gluon condensate\cite{GrSch} $K=(0.36 \pm
0.02\,\mbox{GeV})^4$ gives $\Lambda=(631 \pm 79)\,$MeV (this
estimation corresponds to the one-loop level with $\nf=3$ active
quarks). This value of the parameter \LQCD agrees fairly well with
its estimation obtained in Subsection~\ref{Sect:Vr}.

\subsection{Inclusive $\tau$ lepton decay}
\label{Sect:Tau}

     In the previous subsections the analytic invariant charge has
been applied to description of some intrinsically nonperturbative
issues of QCD, the congruous estimation of the parameter \LQCD being
obtained. For further verification of the self--consistency of the
model proposed, it is worth proceeding to the hadron processes which
are commonly described within the perturbative approach.
Investigation of such processes at relatively small energies  is of a
primary interest here. Among them, the inclusive $\tau$~lepton decay
is a most sensitive to the infrared behavior of the strong running
coupling.  Thus, we turn now to this hadron process, restricting
ourselves at this stage to the one-loop level (see
Ref.~\refcite{PRD2} for details).

     The experimentally measurable quantity here is the inclusive
semileptonic branching ratio
\begin{equation}
\label{RTauDef}
R_{\tau} = \frac{\Gamma(\tau^{-} \to \mbox{hadrons}^{-}\, \nu_{\tau})}
{\Gamma(\tau^{-} \to e^{-}\, \bar\nu_{e}\, \nu_{\tau})}.
\end{equation}
One can split this ratio into three parts, namely
$R_{\tau}=\RTau{V}+\RTau{A}+\RTau{S}$. The terms $\RTau{V}$ and
$\RTau{A}$ account the contributions to Eq.~(\ref{RTauDef}) of the
decay modes with the light quarks only, and they correspond to the
vector (V) and axial--vector~(A) quark currents, respectively. The
accuracy of the experimental measurement of these terms is several
times higher than the accuracy of the strange width ratio~$\RTau{S}$,
which accounts the $s$~quark contribution to Eq.~(\ref{RTauDef}).
Thus, let us proceed with the nonstrange part of the $R_{\tau}$~ratio
associated with the vector quark currents (see
Refs.~\refcite{BDP,BNP,Pi1} and~\refcite{APTTau} for detailed
description of this issue):
\begin{equation}
\label{RTauVDef}
\RTau{V} = \frac{\Nc}{2}\,|\Vud|^2 \Sew \, \left(1 + \dQCD \right).
\end{equation}
In this equation $\Nc=3$ is the number of quark colors,
$|\Vud|=0.9734 \pm 0.0008$ denotes the Cabibbo--Kobayashi--Maskawa
matrix element,\cite{PDG} $\Sew = 1.0194 \pm 0.0040$ is the
electroweak factor,\cite{BNP,EWF} and $\dQCD$ is the strong
correction. A recent experimental measurement of the
ratio~(\ref{RTauVDef}) by ALEPH Collaboration gave\cite{ALEPH}
$\RTau{V}=1.775 \pm 0.017$.

     In the framework of the approach in hand the one-loop QCD
correction  is determined by the integral\cite{PRD2}
\begin{equation}
\label{TauCorDef}
\dQCD = \frac{2}{\pi}\int\limits_{4m^2}^{M_{\tau}^2}
\frac{d s}{M_{\tau}^2} \left(1 - \frac{s}{M_{\tau}^2}\right)^{\! 2}
\left(1 + 2 \frac{s}{M_{\tau}^2}\right) \hal{(1)}{an}(s),
\end{equation}
where $\hal{(1)}{an}(s)$ is the analytic running coupling in the
timelike region~(\ref{AICTL}),
$M_{\tau}=(1776.99_{-0.26}^{+0.29})\,$MeV denotes the $\tau$~lepton
mass, and $m=(4.0 \pm 1.5)\,$MeV stands for the light quark mass\cite{PDG}
at a normalization scale of~$2\,$GeV.  It is worth noting here
that there is no need to involve the contour integration in
Eq.~(\ref{TauCorDef}), since analytic running coupling~(\ref{AICTL})
contains no unphysical singularities in the region~$s>0$. In other
words, the integration in Eq.~(\ref{TauCorDef}) can be performed in a
straightforward way. Introducing the notations $x = s/M^2_{\tau}$,
$x_0 = 4 m^2/M_{\tau}^2$, and $c_0 = x_0(2 - 2 x_0^2 + x_0^3)$, one
can rewrite Eq.~(\ref{TauCorDef}) in a more convenient form
\begin{equation}
\dQCD = \frac{1}{\pi}\Bigl[\hal{(1)}{an}(M_{\tau}^2) -
c_0 \hal{(1)}{an}(4 m^2)\Bigr] +
\frac{4}{\beta_0} \int\limits_{x_0}^{1}
\left(2 - 2x^2 + x^3\right) \ro{1}\!
\left(x\frac{M_{\tau}^2}{\Lambda^2}\right)\, d x,
\end{equation}
where the spectral density $\ro{1}(\sigma)$ is defined in
Eq.~(\ref{SpDns1L}).

     For the value $\RTau{V}$ given above\cite{ALEPH} the estimation
$\Lambda^{\!(\nfs=2)}=(561 \pm 69)\,$MeV has been obtained for two
active quarks. The uncertainty here is due to the errors in the
values of $\RTau{V}$, $|\Vud|$, $\Sew$, $m$, and $M_{\tau}$. However,
in order to collate this result with the previous estimations  of the
parameter $\Lambda_{\mbox{\tiny QCD}}$, one has to continue it to the
region of three active quarks (the $s$~quark mass\cite{PDG} $m_s =
(117.5 \pm 37.5)\,$MeV has been employed here). This gives the value
$\Lambda^{\!(\nfs=3)}=(524 \pm 68)\,$MeV, which perfectly agrees with
all the estimations of \LQCD in the framework of the approach in
hand.

\subsection{$e^{+}e^{-}$ annihilation into hadrons}
\label{Sect:EpEm}

     It is worthwhile to consider one more strong interaction
process, namely,  the electron--positron annihilation into hadrons.
The experimentally measurable  quantity here is the ratio of two
cross--sections
\begin{equation}
\label{RGen}
R(s) = \frac{\sigma(e^{+}e^{-} \to \mbox{hadrons})}
{\sigma(e^{+}e^{-} \to \mu^{+}\mu^{-})}.
\end{equation}
Unlike the inclusive semileptonic branching ratio~(\ref{RTauDef}),
this equation contains the explicit dependence on the
center--of--mass energy of the process. In the framework of the
approach in hand, at the one-loop level the ratio~$R(s)$ can be
represented in the following form (see, e.g.,
Refs.~\refcite{AGZ} and~\refcite{APTEpEm}):
\begin{equation}
R(s) = \Nc \sum_{f=1}^{\nfs} Q_f^2 \left[1 + \frac{1}{\pi}
\hal{(1)}{an}(s)\right].
\end{equation}
In this equation $\Nc=3$ is the number of colors, $Q_f$ denotes the
charge of the $f$-th quark, and $\hal{(1)}{an}(s)$ is the analytic
running coupling in the timelike region~(\ref{AICTL}).

     Recent experimental measurement of the $R(s)$ by CLEO
Collaboration at $\sqrt{s_0}=10.52\,$GeV gave\cite{CLEO} $R(s_0)=3.56
\pm 0.01\,$(stat.). With this value the estimation of the parameter
\LQCD  for $\nf=4$ active quarks is $\Lambda^{\!(\nfs=4)}  = (393 \pm
65)\,$MeV. It is worth noting that this estimation turns out to be
rather rough due to a considerable systematic errors of the
experimental data at small energies. Nevertheless, the continuation
of the obtained result to the three active quarks region gives
$\Lambda^{\!(\nfs=3)}=(490 \pm 76)\,$MeV, that agrees fairly well
with all the previous estimations of the
parameter~$\Lambda_{\mbox{\tiny QCD}}$.

\medskip

     Thus, the applications of the model developed to description of
diverse strong interaction processes ultimately lead to congruous
estimation of the parameter~$\Lambda_{\mbox{\tiny QCD}}$. Namely, at
the one-loop level with $\nf = 3$ active quarks $\Lambda = (557 \pm
36)\,$MeV. Apparently, this testifies that the analytic invariant
charge substantially incorporates, in a consistent way, both
perturbative and intrinsically nonperturbative aspects of Quantum
Chromodynamics.

% SECTION 5

\section{Conclusions}
\label{Sect:Concl}

     Let us summarize the basic properties of the model for the QCD
analytic invariant charge, considered in this review,  and the main
results obtained in this field.

     The proposed way of involving the analyticity requirement into
the renormalization group method results in qualitatively new
features of the strong running coupling. Namely, analytic invariant
charge has no unphysical singularities, and it incorporates the
ultraviolet asymptotic freedom with infrared enhancement in a single
expression. Furthermore, additional parameters are not introduced
into the theory. The consistency of the model  developed with the
general definition of the QCD invariant charge is proved.

     The detailed investigation of the one-loop analytic running
coupling and the relevant $\beta$~function is performed. The latter
is proved to coincide with its perturbative analog at small values of
invariant charge. At the same time, this $\beta$~function possesses
the asymptotic behavior corresponding to the infrared enhancement of
the strong running coupling. The renormalization invariance of the
analytic running coupling is shown explicitly.

     The analytic invariant charge is derived and studied at the
higher loop levels also. It is proved that the higher loop
corrections do not affect the AIC qualitatively. In particular, the
analytic running coupling has the universal behavior both in
ultraviolet  and infrared regions at any loop level, and it possesses
good higher loop and scheme stability.

     The analytic invariant charge is extended to the timelike
region. It is shown to have asymmetrical behavior in the
intermediate-- and low--energy domains of the spacelike and timelike
regions. This difference turns out to be rather substantial and
should be taken into account when one handles the experimental data.
The obtained result confirms hypothesis due to Schwinger concerning
the relation between the $\beta$~function and the relevant spectral
density.

     The developed model is applied to description of diverse strong
interaction processes both of perturbative and intrinsically
nonperturbative nature. The derived static quark--antiquark potential
is proved to be confining at large distances. At the same time, it
has the standard perturbative behavior at small distances. The
analytic invariant charge reproduces explicitly the conformal
inversion symmetry related to the size distribution of instantons.
The developed model is also applied to description of gluon
condensate, inclusive $\tau$~lepton decay and electron--positron
annihilation into hadrons. The congruity of the estimated  values of
the parameter \LQCD testifies to the self--consistency of the
approach developed.

\section*{Acknowledgments}

     The author expresses heartfelt gratitude to D.V.~Shirkov and
I.L.~Solovtsov for the scientific guidance and a very fruitful
collaboration. The author is grateful to F.~Schrempp for  valuable
discussions and stimulating comments and to G.S.~Bali for supplying
the relevant lattice data and useful remarks. It is a pleasant duty
for the author to thank also A.I.~Alekseev, B.A.~Arbuzov,
V.V.~Belokurov, N.~Brambilla, R.M.~Corless, Yu.L.~Dokshitzer, H.M.~Fried,
G.~Grunberg, V.V.~Kiselev, S.V.~Mikhailov, B.~Pire, A.A.~Pivovarov,
C.~Roiesnel, A.V.~Sidorov, L.~von~Smekal, O.P.~Solovtsova, M.~Teper,
and O.V.~Teryaev for useful discussions and stimulating comments. The
author thanks the CPHT \'Ecole Polytechnique for very warm
hospitality. The partial support of RFBR (grants 02--01--00601,
04--02--81025, and 00--15--96691) is appreciated.

% APPENDIX

\appendix

\section{Perturbative QCD Running Coupling}
\label{Sect:RCPert}

     The QCD invariant charge $g^2(\mu^2)$ is the solution to
the renormalization group equation
\begin{equation}
\label{RGEqn}
\frac{d\,\ln \bigl[g^2(\mu^2)\bigr]}{d\,\ln\mu^2} =
\beta\Bigl(g(\mu^2)\Bigr).
\end{equation}
In the framework of perturbative approach, assuming the running
coupling being sufficiently small, one can approximate the
$\beta$~function on  the right-hand side of this equation by the
power series
\begin{equation}
\beta\Bigl(g(\mu^2)\Bigr) = - \left\{
\beta_{0}\left[\frac{g^2(\mu^2)}{16 \pi^2}\right] +
\beta_{1}\left[\frac{g^2(\mu^2)}{16 \pi^2}\right]^2 +
\beta_{2}\left[\frac{g^2(\mu^2)}{16 \pi^2}\right]^3 +
\cdots \right\},
\end{equation}
where for the SU(3) gauge group $\beta_0 = 11 - 2 \nf / 3,\,$
$\beta_1 = 102 - 38 \nf / 3,\,$ $\beta_2 = 2857/2 - 5033\nf/18 +
32\nf^2/54,\,$ and $\nf$ denotes  the number of active quarks. The
coefficients $\bz$ and $\bu$ are scheme independent, while the value
given for the three-loop coefficient $\beta_2$ is calculated in the
\MSbar scheme.\cite{TVZ} Let us rewrite RG equation~(\ref{RGEqn}) at
the $\ell$-loop level as follows:
\begin{equation}
\label{RGEqnPert}
\frac{d\,\ln\bigl[\tal{(\ell)}{s}(\mu^2)\bigr]}{d\,\ln \mu^2} = -
\sum_{j=0}^{\ell-1} B_j
\left[\tal{(\ell)}{s}(\mu^2)\right]^{j+1}, \qquad
B_j=\frac{\beta_j}{\beta_0^{j+1}},
\end{equation}
where $\al{}{s}(\mu^2) = g^2(\mu^2)/(4\pi)$ and $\tal{}{}(\mu^2)
= \alpha(\mu^2)\beta_0/(4\pi)$.

     At the one-loop level Eq.~(\ref{RGEqnPert}) reads as
\begin{equation}
\frac{d\,\ln\bigl[\tal{(1)}{s}(\mu^2)\bigr]}{d\,\ln \mu^2} =
- \tal{(1)}{s}(\mu^2).
\end{equation}
Integrating this equation in finite terms, one gets
\begin{equation}
\frac{1}{\tal{(1)}{s}(q^2)} - \frac{1}{\tal{(1)}{s}(q_0^2)} =
\ln\!\left(\frac{q^2}{q_0^2}\right).
\end{equation}
It is convenient to introduce here the parameter $\Lambda$ of the
dimension of mass, which absorbs all the dependence on the
normalization point~$q_0^2$:
\begin{equation}
\label{LDef1L}
\Lambda\ind{2}{(1)} = q_0^2\,\exp\!\left[- \frac{4 \pi}{\bz}
\frac{1}{\al{(1)}{s}(q_0^2)}\right].
\end{equation}
In this case, the solution to the RG equation~(\ref{RGEqnPert}) takes
the well-known form\footnote{It is worth mentioning here that the
relation~(\ref{LDef1L}) ensures the renormalization invariance of the
solution to the RG equation~(\ref{RCPert1L}).}
\begin{equation}
\label{RCPert1L}
\al{(1)}{s}(q^2) = \frac{4 \pi}{\bz} \, \frac{1}{\ln z}, \qquad
z = \frac{q^2}{\Lambda^2}.
\end{equation}
Let us note here that this expression for the one-loop perturbative
running coupling has the unphysical singularity (the so-called Landau
pole) at the point~$q^2=\Lambda^2$.

     At the two-loop level Eq.~(\ref{RGEqnPert}) acquires the form
\begin{equation}
\label{RGP1L2}
\frac{d\,\ln\bigl[\tal{(2)}{s}(\mu^2)\bigr]}{d\,\ln \mu^2} =
- \tal{(2)}{s}(\mu^2) - B_1 \left[\tal{(2)}{s}(\mu^2)\right]^2.
\end{equation}
Integrating this equation likewise the previous case, one arrives at
the transcendental relation,\footnote{Certainly, Eq.~(\ref{RGP1L2})
has also solutions different from Eq.~(\ref{RGP2L2}). Nevertheless,
all the ambiguity here can be eliminated by redefinition of the
parameter~$\Lambda$ on the right-hand side of Eq.~(\ref{RGP2L2}).}
which determines the two-loop perturbative running coupling
\begin{equation}
\label{RGP2L2}
\frac{1}{\tal{(2)}{s}(q^2)} - B_1 \ln\left[1 +
\frac{1}{B_1 \tal{(2)}{s}(q^2)}\right] =
\ln\!\left(\frac{q^2}{\Lambda^2}\right).
\end{equation}
The parameter $\Lambda$ in this formula differs from the one-loop
one~(\ref{LDef1L})
\begin{equation}
\label{LDef2L}
\Lambda\ind{2}{(2)} = \mu^2 \exp\!\left\{- \frac{4 \pi}{\bz}
\frac{1}{\al{(2)}{s}(\mu^2)} +
B_1\ln\!\left[1 + \frac{1}{B_1}\frac{4 \pi}{\bz}
\frac{1}{\al{(2)}{s}(\mu^2)} \right]\right\}.
\end{equation}
The explicit solution to Eq.~(\ref{RGP2L2}) can be written down in
terms of the so-called Lambert~$W$ function\cite{GGKM}
(see~\ref{Sect:LambW}):
\begin{equation}
\label{RCPert2LEx}
\al{(2)}{ex}(q^2) = - \frac{4 \pi}{\bz} \frac{1}{B_1}
\frac{1}{1 + W_{k}\left\{-\exp\!\left[-\!\left(1 + B_1^{-1}\ln z
\right)\right]\right\}},
\end{equation}
where $z=q^2/\Lambda^2$ and $B_1=\beta_1/\beta_0^2$. Here the branch
index  of the Lambert~$W$ function may take the values $k=0$ or
$k=-1$. But only the latter one satisfies the asymptotic freedom
condition. It is worthwhile to note that two-loop perturbative
invariant charge~(\ref{RCPert2LEx}) also  has unphysical
singularities, namely, the pole at $q^2=\Lambda^2$ and the cut
$0<q^2\le\Lambda^2$. The exact solution~(\ref{RCPert2LEx}) is widely
used along with the iterative solution to Eq.~(\ref{RGP2L2})
\begin{equation}
\label{RCPert2LIt}
\al{(2)}{it}(q^2) = \frac{4 \pi}{\bz}
\frac{1}{\ln z + B_1 \ln\!\left(1 + B_1^{-1} \ln z\right)}
\end{equation}
and its approximate form (see, e.g., Ref.~\refcite{Ynd}):
\begin{equation}
\al{(2)}{ap}(q^2) = \frac{4 \pi}{\bz} \frac{1}{\ln z}
\left[1 - B_1 \frac{\ln (\ln z)}{\ln z}\right].
\end{equation}

     Since the account of the higher-order corrections leads to
substantial technical complications, we present only the explicit
expression for the three-loop running coupling here (see, e.g.,
Ref.~\refcite{PDG}):
\begin{eqnarray}
\al{(3)}{s}(q^2) &=& \frac{4 \pi}{\bz} \frac{1}{\ln z}
\left\{1 - \frac{B_1}{\ln z} \ln (\ln z) +
\frac{B_1^2}{(\ln z)^2} \biggl[ \left(\ln(\ln z)\right)^2
- \ln(\ln z) + \frac{B_2}{B_1^2} - 1\biggr] \right\}.
\nonumber \\ && \label{RCPert3L}
\end{eqnarray}
The calculation of the four-loop $\beta$~function
coefficient~$\beta_3$ in the \MSbar scheme has been performed in
Ref.~\refcite{RVL}. For the relevant four-loop perturbative running
coupling see, e.g., Ref.~\refcite{Alek01}.

\section{The Lambert $W$ Function}
\label{Sect:LambW}

     For the representation of the analytic running
coupling~(\ref{AIC1L}) in the renorminvariant form and for the
investigation of the corresponding $\beta$~function~(\ref{BetaAn}) it
proves to be convenient to use the so-called Lambert~$W$ function. As
long ago as the middle of $18$th century this function is being
employed in diverse physical problems.\cite{Corl} In contemporary
QCD studies the interest to this function arose just a several years
ago. Namely, it was revealed\cite{GGKM} that the explicit
solution to the perturbative RG equation for the invariant charge at
the two-loop level can be expressed in terms of the Lambert~$W$
function (see also~\ref{Sect:RCPert}).

\begin{figure}[t]
\centerline{\psfig{file=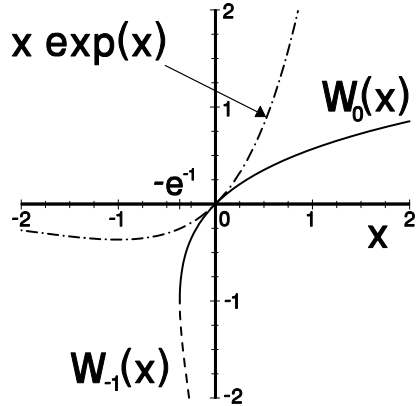,width=55mm}}
\vspace*{8pt}
\caption{The function $x\,e^x$ (dot-dashed curve) and two real
branches of the Lambert~$W$ function~(\ref{LambWDef}): $W_{0}(x)$
(solid curve) and $W_{-1}(x)$ (dashed curve).}
\label{Plot:LambertW}
\end{figure}

     The Lambert~$W$ function is defined as a many--valued function
$W_k(x)$, which satisfies the equation
\begin{equation}
\label{LambWDef}
W_{k}(x)\,\exp\Bigl[W_{k}(x)\Bigr] = x,
\end{equation}
where $k$ denotes the branch index of this function. Only two real
branches of the Lambert~$W$ function, the principal  branch $W_0(x)$
and the branch $W_{-1}(x)$ (see Fig.~\ref{Plot:LambertW}), will be
used in our consideration. The other branches of this function take
an imaginary values. One can show that for the branches $W_0(x)$ and
$W_{-1}(x)$ the following expansions hold:
\begin{eqnarray}
\label{WSeries1}
W_{0}(\varepsilon) &=& \varepsilon - \varepsilon^2 +
{\cal O}(\varepsilon^3),
\quad \varepsilon \to 0, \\
W_{-1}(-\varepsilon) &=& \ln \varepsilon +
{\cal O}\left(\ln |\ln\varepsilon|\right),
\quad \varepsilon \to 0_{+}, \\
W_{0}\left(-\frac{1}{e} + \varepsilon \right) &=& -1 + \sqrt{2 e
\varepsilon} + {\cal O}(\varepsilon), \quad \varepsilon \to 0_{+},\\
\label{WSeries2}
W_{-1}\left(-\frac{1}{e} + \varepsilon \right) &=& -1 - \sqrt{2 e
\varepsilon} + {\cal O}(\varepsilon), \quad \varepsilon \to 0_{+},
\end{eqnarray}
where $e=2.71828\ldots$ denotes the base of the natural logarithm.
Details concerning the properties of the Lambert $W$ function can be
found in Ref.~\refcite{Corl}.

     In order to represent the analytic running
coupling~(\ref{AIC1L}) in a renorminvariant form and to derive the
corresponding $\beta$~function~(\ref{BetaAn}) one has to solve the
equation (see Subsection~\ref{Sect:AIC1L} and
Refs.~\refcite{MPLA2,MPLA1})
\begin{equation}
\label{EqnGen}
\frac{z-1}{z\,\ln z} = a,\qquad z>0,\qquad a>0
\end{equation}
for the variable $z$. Let us multiply this equation through by the
factor $a^{-1}\,\ln z$
\begin{equation}
\label{Eqn1}
\ln z = \frac{b}{z} - b,\qquad z>0,
\end{equation}
where $b=-1/a$. Obviously, the equation obtained has a trivial
solution $z=1$, which does not satisfy the initial Eq.~(\ref{EqnGen})
when $a \neq 1$, that is a consequence of the multiplication of the
latter by~$\ln z$. Therefore, when solving Eq.~(\ref{Eqn1}) for the
variable~$z$ one has to discard its trivial solution $z=1$. Next, let
us represent Eq.~(\ref{Eqn1}) in the form
\begin{equation}
\label{Eqn2}
\frac{b}{z}\,\exp\!\left(\frac{b}{z}\right) = b\,e^{b}.
\end{equation}
Taking into account the definition~(\ref{LambWDef}), solution to
Eq.~(\ref{Eqn2}) can be expressed in terms of the Lambert~$W$ function
\begin{equation}
\label{Sol}
\frac{b}{z} = W_{k} \left(b\,e^b\right).
\end{equation}
The branch index $k$ of the $W$ function will be specified below.

\begin{figure}
\centerline{\psfig{file=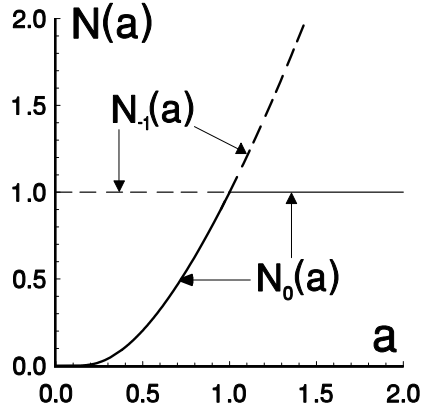,width=55mm}}
\vspace*{8pt}
\caption{The function $N(a)$ (boldface curve), and the functions
$N_{0}(a)$ (solid curve) and $N_{-1}(a)$ (dashed curve) (see
Eq.~(\protect\ref{LambW:NDef})).}
\label{Plot:N}
\end{figure}

     In the physically relevant range $a>0$ the argument of the
Lambert~$W$ function in Eq.~(\ref{Sol}) takes the values $-1/e\le
b\,e^{b} < 0$. The only two real branches, the principle branch
$W_0(x)$ and the branch $W_{-1}(x)$ (see Fig.~\ref{Plot:LambertW})
correspond to this interval. Here one has to handle carefully with
the interchange between these branches at the point $x=-1/e$
(this corresponds to the value $a=1$). Thus, the nontrivial solution
to Eq.~(\ref{Eqn1}) for the variable $z$ is
\begin{equation}
\frac{1}{z} = \left\{
\begin{array}{lrl}
b^{-1}\, W_{-1}\!\left(b\,e^b\right), \quad & -1 \le b < 0,& \\[2.5mm]
b^{-1}\, W_{0}\!\left(b\,e^b\right),        &        b < -1& \\
\end{array}
\right.
\end{equation}
(another choice of branches will be considered below). Therefore, the
solution to Eq.~(\ref{EqnGen}) we are interested in can be written in
the form
\begin{equation}
z = \frac{1}{N(a)},
\end{equation}
where the function $N(a)$ (see Fig.~\ref{Plot:N}) is defined by
\begin{equation}
\label{LambW:NDef}
N(a) = \left\{
\begin{array}{ll}
N_{0}(a),        & 0 < a \le 1, \\[2.5mm]
N_{-1}(a), \quad & 1 < a,       \\
\end{array}
\right. \qquad
N_{k}(a) = -a\, W_{k}\!
\left[-\frac{1}{a}\,\exp\!\left(-\frac{1}{a}\right)\right].
\end{equation}

     As follows from the definition~(\ref{LambWDef}), the function
$xe^x$ and the Lambert~$W$ function are mutually inverse. Here one
has to distinguish precisely the branches $W_0$ and $W_{-1}$ (see
Fig.~\ref{Plot:LambertW}). From this figure it follows directly that
for $x > -1$ the function $x e^x$ ``corresponds'' to the branch
$W_0$, and for $x \le -1$ the function $x e^x$ ``corresponds'' to the
branch $W_{-1}$. Let us introduce for convenience the notations of
the inverse functions for these branches
\begin{equation}
x\,e^x = \left\{
\begin{array}{ll}
\left(W_{0}\right)^{-1}\!(x),        & x > -1,   \\[2.5mm]
\left(W_{-1}\right)^{-1}\!(x), \quad & x \le -1. \\
\end{array}
\right.
\end{equation}
Now the function $N(a)$ in Eq.~(\ref{LambW:NDef}) can be written in a
more compact form
\begin{equation}
N(a) = \left\{
\begin{array}{ll}
-a\,W_{0}\!\left[\left(W_{-1}\right)^{-1}\!\left(-a^{-1}\right)\right], \quad
& 0 < a \le 1, \\[2.5mm]
-a\,W_{-1}\!\left[\left(W_{0}\right)^{-1}\!\left(-a^{-1}\right)\right],
& 1 < a. \\
\end{array}
\right.
\end{equation}
It is worth noting also that another choice of the branches of the
function~$N(a)$
\begin{equation}
\widetilde{N}(a) = \left\{
\begin{array}{ll}
N_{-1}(a), \quad & 0 < a \le 1, \\[2.5mm]
N_{0}(a),        & 1 < a        \\
\end{array}
\right.
\end{equation}
leads to the trivial solution of Eq.~(\ref{Eqn1}) (in this case
$\widetilde{N}(a) \equiv 1$). Therefore, the solution
$z=1/\widetilde{N}(a)$ does not satisfy Eq.~(\ref{EqnGen}) when $a\neq 1$.

     Let us outline briefly the basic properties of the
function~$N(a)$ introduced in Eq.~(\ref{LambW:NDef}). In the physical
range $a>0$ it is a non-negative, monotonously increasing function
(see Fig.~\ref{Plot:N}). With the help of the
series~(\ref{WSeries1})--(\ref{WSeries2}) one can show that for the
function $N(a)$ the following expansions hold:
\begin{eqnarray}
\label{NSeries1}
N(a\to 0_{+}) &=& \exp\!\left(-\frac{1}{a}\right)
\left\{1 + {\cal O}\!\left[\exp\!\left(-\frac{1}{a}\right)\right]\right\}, \\
N(1+\varepsilon) &=& 1 + 2\varepsilon + {\cal O}(\varepsilon^2), \quad
\varepsilon \to 0, \\
\label{NSeries2}
N(a\to\infty) &=& a\ln a\left\{
1+{\cal O}\!\left[\frac{\ln(\ln a)}{\ln a}
\right]\right\}.
\end{eqnarray}

\section{Spectral Functions}
\label{Sect:SF}

     The spectral function~$\Ro{\ell}$ at the $\ell$-loop level has
the following form (see equation~(\ref{SpFunDef})):
\begin{equation}
\label{SpFunDef1}
\Ro{\ell} = \frac{1}{2 \pi i}
\lim_{\varepsilon \to 0_{+}}
\sum_{j=0}^{\ell-1} \frac{\beta_j}{\beta_0^{j+1}}
\biggl\{\Bigl[\tal{(\ell)}{s}(-\sigma-i\varepsilon)\Bigr]^{j+1}
- \Bigl[\tal{(\ell)}{s}(-\sigma+i\varepsilon)\Bigr]^{j+1}\biggr\}.
\end{equation}
Here $\al{(\ell)}{s}(\mu^2)$ is the $\ell$-loop perturbative running
coupling and $\beta_j$ denotes the $\beta$~function expansion
coefficient (see~\ref{Sect:RCPert}). It turns out to be  convenient
to rearrange equation~(\ref{SpFunDef1}):
\begin{equation}
\label{SpFunDef2}
\Ro{\ell} = \sum_{j=0}^{\ell-1} B_j\, \varrho\ind{(\ell)}{j+1}(\sigma),
\end{equation}
where
\begin{equation}
\varrho\ind{(\ell)}{j+1}(\sigma) = \frac{1}{2\pi i}\,
\lim_{\varepsilon \to 0_{+}} \biggl\{
\Bigl[\tal{(\ell)}{s}(-\sigma-i\varepsilon)\Bigr]^{j+1} -
\Bigl[\tal{(\ell)}{s}(-\sigma+i\varepsilon)\Bigr]^{j+1}
\biggr\},
\end{equation}
and $B_j=\beta_j/\beta_0^{j+1}$. It is worthwhile to quote also
several commonly used relations:
\begin{equation}
\lim_{\varepsilon \to 0_{+}} \ln (-\sigma \pm i\varepsilon) =
\ln \sigma \pm i\pi, \qquad \sigma > 0
\end{equation}
and
\begin{equation}
\label{ComplLog}
\ln (a \pm i b) = \frac{1}{2} \ln (a^2 + b^2) \pm
i\left[\frac{\pi}{2} - \arctan\!\left(\frac{a}{b}\right)\right],
\qquad b > 0.
\end{equation}
Here it is assumed that $-\pi/2\leq\arctan (x)\leq\pi/2$.

     The one-loop perturbative running coupling is determined by
Eq.~(\ref{RCPert1L}). In this case the spectral
function~(\ref{SpFunDef1}) takes a simple form
\begin{equation}
\label{SpFun1L}
\Ro{1} = \varrho\ind{(1)}{1}(\sigma) =
\frac{1}{\ln^2 (\sigma/\Lambda^2) + \pi^2}.
\end{equation}

     For the construction of the two-loop spectral function we shall
use the expression~(\ref{RCPert2LIt}),\footnote{It has been shown in
Ref.~\refcite{APTRev} that the spectral functions constructed by making
use of Eqs.~(\ref{RCPert2LIt}) and~(\ref{RCPert3L}) practically lead
to the same result as the spectral functions corresponding  to the
exact (numerical) solutions to the RG equation~(\ref{RGEqnPert}) at
the higher-loop levels.} which gives
\begin{equation}
\label{SpFun2L}
\Ro{2} = \varrho\ind{(2)}{1}(\sigma) +
B_1 \varrho\ind{(2)}{2}(\sigma),
\end{equation}
where
\begin{eqnarray}
\varrho\ind{(2)}{1}(\sigma) &=&
\frac{\Ctwo}{y^2 \left[\Atwo\right]^2 + \pi^2 \left[\Ctwo\right]^2}, \\
\varrho\ind{(2)}{2}(\sigma) &=&
\frac{2y \Atwo \Ctwo}{\left\{y^2 \left[\Atwo\right]^2 +
\pi^2 \left[\Ctwo\right]^2\right\}^2},
\end{eqnarray}
with the two-loop functions $\Atwo$ and $\Ctwo$ defined by
\begin{eqnarray}
\Atwo &=& 1 + \frac{B_1}{y}\left\{\frac{1}{2}\ln\!\Bigl[
(y+B_1)^2 + \pi^2 \Bigr] - \ln B_1 \right\}, \\
\Ctwo &=& 1 + B_1 \left[\frac{1}{2} - \frac{1}{\pi} \arctan\!\left(
\frac{y+B_1}{\pi}\right)\right].
\end{eqnarray}
Here and further we use the notation $y=\ln (\sigma /\Lambda^2)$.

     By making use of the perturbative running
coupling~(\ref{RCPert3L}), the following result was obtained for the
three-loop spectral function~(\ref{SpFunDef1}):
\begin{equation}
\label{SpFun3L}
\Ro{3} = \varrho\ind{(3)}{1}(\sigma) +
B_1 \varrho\ind{(3)}{2}(\sigma) + B_2 \varrho\ind{(3)}{3}(\sigma),
\end{equation}
where
\begin{eqnarray}
\varrho\ind{(3)}{1}(\sigma) &=& \frac{\Cthree}{(y^2 + \pi^2)^3}, \\
\varrho\ind{(3)}{2}(\sigma) &=& \frac{2 \Athree \Cthree}{(y^2 + \pi^2)^6}, \\
\varrho\ind{(3)}{3}(\sigma) &=&
\frac{\Cthree \left\{3\left[\Athree\right]^2 -
\pi^2\left[\Cthree\right]^2\right\}}
{(y^2 + \pi^2)^9},
\end{eqnarray}
and the three-loop functions $\Athree$ and $\Cthree$ are
\begin{eqnarray}
\Athree &=& y (y^2 - 3\pi^2) R(y) - \pi^2 (\pi^2 - 3y^2) J(y), \\
\Cthree &=& y (3\pi^2 - y^2) J(y) - (3y^2 - \pi^2) R(y), \\
R(y) &=& y^2 - \pi^2 - B_1 (y a - \pi^2 c) +
B_1^2 (a^2 - \pi^2 c^2 - a - 1) + B_2, \\
J(y) &=& 2y -B_1 (y c + a) + B_1^2 c (2 a - 1), \\
a &=& \frac{1}{2} \ln (y^2 + \pi^2), \qquad
c = \frac{1}{2} - \frac{1}{\pi} \arctan\!\left(\frac{y}{\pi}\right).
\end{eqnarray}

\begin{figure}
\centerline{\psfig{file=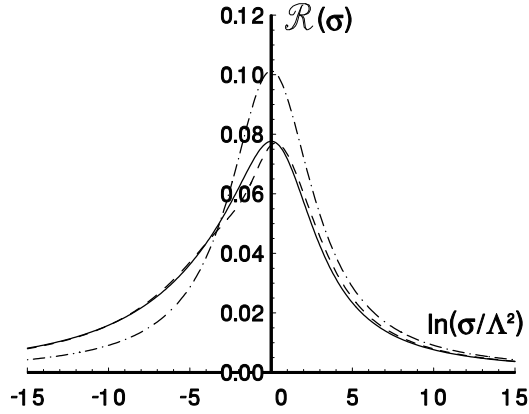,width=70mm}}
\vspace*{8pt}
\caption{The spectral functions $\Ro{\ell}$ defined by
Eq.~(\ref{SpFunDef1}). The one-, two-, and three-loop levels are
shown by the dot-dashed, solid, and dashed curves, respectively.}
\label{Plot:SpFun}
\end{figure}

     The plots of the spectral functions~(\ref{SpFun1L}),
(\ref{SpFun2L}), and (\ref{SpFun3L}) are presented in
Figure~\ref{Plot:SpFun}. It is interesting to note that $\Ro{\ell}
\to \Ro{1}$, when $\sigma \to \infty$, as well as when $\sigma \to 0$,
irrespective of the loop level. It is worth mentioning also that the
relation  $\max |\varrho\ind{(\ell)}{j}(\sigma)| \gg
\max|\varrho\ind{(\ell)}{j+1}(\sigma)|$ holds at any loop level. In
particular,  this implies that the contribution of the every
subsequent term on the right-hand side of Eq.~(\ref{RGEqnAnHL}) is
substantially suppressed by the contributions of the preceding ones.
In turn, this results in a good higher loop and scheme stability of
the approach in hand.

\section{The Expansion Coefficients of the \\
         Quark--Antiquark Potential}
\label{Sect:VrExp}

     In this appendix the explicit expressions for the expansion
coefficients of the static quark--antiquark potential generated by
the analytic invariant charge~(\ref{VrReg}) are gathered in a concise
form. By making use of Eqs.~(\ref{uwDef}) and~(\ref{vDef}) we find

\begin{eqnarray}
u_0 &=&  1, \\
u_1 &=&  -\gamma, \\
u_2 &=& \frac{\pi^2}{12} + \gamma^2, \\
u_3 &=& -2 \zeta(3) - \frac{\pi^2}{4}\gamma - \gamma^3, \\
u_4 &=& \frac{19}{240} \pi^4 + 8 \zeta(3) \gamma +
        \frac{\pi^2}{2}\gamma^2 + \gamma^4, \\
\omega_0 &=& \frac{1}{2}, \\
\omega_1 &=& \frac{3}{4} - \frac{\gamma}{2}, \\
\omega_2 &=& \frac{7}{4} + \frac{\pi^2}{24} - \frac{3}{2} \gamma +
             \frac{1}{2}\gamma^2, \\
\omega_3 &=& \frac{45}{8} + \frac{3}{16}\pi^2 - \zeta(3) -
\frac{\gamma}{4}\!\left[\frac{\pi^2}{2}+21\right] + \frac{9}{4}\gamma^2
-\frac{1}{2}\gamma^3, \\
\omega_4 &=& \frac{93}{4} + \frac{\pi^2}{8}\!\left[\frac{19}{60}\pi^2 +
7 \right]- 6 \zeta(3) + \gamma\left[4\zeta(3) - \frac{45}{2} -
\frac{3}{4}\pi^2\right]
\nonumber \nopagebreak \\
&& + \frac{\gamma^2}{2}\!\left[\frac{\pi^2}{2}+21\right]
-3 \gamma^3 + \frac{1}{2}\gamma^4,
\end{eqnarray}
where $\gamma \simeq 0.57721...$ denotes the Euler's constant, and
$\zeta(x)$ is the Riemann's Zeta function\cite{RG} $(\zeta(3) \simeq
1.20206...)$. Applying the ``regularization'' procedure specified in
Subsection~\ref{Sect:Vr} one obtains (see Eqs.~(\ref{VrReg})
and~(\ref{Vr}))
\begin{eqnarray}
\left[\frac{1}{R\,\ln R} \right]_\Reg &=& \frac{1}{R\,\ln R} - \frac{1}{R-1}, \\
\left[\frac{1}{R\,(\ln R)^2} \right]_\Reg &=& \frac{1}{R\,(\ln R)^2} -
  \frac{1}{(R-1)^2}, \\
\left[\frac{1}{R\,(\ln R)^3} \right]_\Reg &=& \frac{1}{R\,(\ln R)^3} -
\frac{1}{2}\frac{1}{(R-1)^2} - \frac{1}{(R-1)^3}, \\
\left[\frac{1}{R\,(\ln R)^4} \right]_\Reg &=& \frac{1}{R\,(\ln R)^4} -
\frac{1}{6}\frac{1}{(R-1)^2} - \frac{1}{(R-1)^3} - \frac{1}{(R-1)^4}, \\
\left[\frac{1}{R\,(\ln R)^5} \right]_\Reg &=& \frac{1}{R\,(\ln R)^5} -
\frac{1}{24}\frac{1}{(R-1)^2} -\frac{7}{12}\frac{1}{(R-1)^3}
\nonumber \nopagebreak \\
&& - \frac{3}{2}\frac{1}{(R-1)^4} - \frac{1}{(R-1)^5}, \\
\left[\frac{R}{\ln R} \right]_\Reg &=& \frac{R}{\ln R} - \frac{1}{R-1}, \\
\left[\frac{R}{(\ln R)^2} \right]_\Reg &=& \frac{R}{(\ln R)^2} -
\frac{2}{R-1} - \frac{1}{(R-1)^2}, \\
\left[\frac{R}{(\ln R)^3} \right]_\Reg &=& \frac{R}{(\ln R)^3} -
\frac{2}{R-1} - \frac{5}{2}\frac{1}{(R-1)^2} - \frac{1}{(R-1)^3}, \\
\left[\frac{R}{(\ln R)^4} \right]_\Reg &=& \frac{R}{(\ln R)^4} -
\frac{4}{3}\frac{1}{R-1} - \frac{19}{6}\frac{1}{(R-1)^2}
\nonumber \nopagebreak \\
&& - \frac{3}{(R-1)^3} - \frac{1}{(R-1)^4}, \\
\left[\frac{R}{(\ln R)^5} \right]_\Reg &=& \frac{R}{(\ln R)^5} -
\frac{2}{3}\frac{1}{R-1} - \frac{65}{24}\frac{1}{(R-1)^2} -
\frac{55}{12}\frac{1}{(R-1)^3} \hspace{25mm}
\nonumber \nopagebreak \\
&& - \frac{7}{2}\frac{1}{(R-1)^4} - \frac{1}{(R-1)^5}.
\end{eqnarray}

% REFERENCES

\end{document}